\documentclass[aps,pre,reprint, superscriptaddress]{revtex4-1}

\usepackage{amsmath}
\usepackage{amsthm}
\usepackage{amsfonts}
\usepackage{amssymb}
\usepackage{mathtools}
\usepackage{graphicx}
\usepackage{hyperref}

\renewcommand{\vec}[1]{\mathbf{#1}}

\begin{document}
	
	\title{Extending the average spectrum method:  Grid points sampling  and density averaging}
	\author{Khaldoon Ghanem} 
	\affiliation{J\"ulich Supercomputer Centre, Forschungszentrum J\"ulich, 52425 J\"ulich, Germany}
	\affiliation{Max-Planck-Institut f\"ur Festk\"orperforschung, 70569 Stuttgart, Germany}
	\author{Erik Koch} 
	\affiliation{J\"ulich Supercomputer Centre, Forschungszentrum J\"ulich, 52425 J\"ulich, Germany}
	\affiliation{JARA High-Performance Computing, 52425 J\"ulich, Germany}
	
	\date{\today}
	
	\begin{abstract}
		Analytic continuation of imaginary time or frequency data to the real axis is a crucial step in extracting dynamical properties from quantum Monte Carlo simulations. 
		The average spectrum method provides an elegant solution by integrating over all non-negative spectra weighted by how well they fit the data. 
		In a recent paper, we found that discretizing the functional integral as in Feynman's path-integrals, does not have a well-defined continuum limit.
		Instead, the limit depends on the discretization grid whose choice may strongly bias the results.
		In this paper, we demonstrate that sampling the grid points, instead of keeping them fixed, also changes the functional integral limit and rather helps to overcome the bias considerably.
		We provide an efficient algorithm for doing the sampling and show how the density  of the grid points acts now as a default model with a significantly reduced biasing effect.
		The remaining bias depends mainly on the width of the grid density, so we go one step further and average also over densities of different widths.
		For a certain class of densities, including Gaussian and exponential ones, this width averaging can be done analytically, eliminating the need to specify this parameter without introducing any computational overhead. 
	\end{abstract}
	
	\maketitle
	
	Quantum Monte Carlo (QMC) simulations have become an indispensable tool for studying quantum many-body systems.
	They often compute Green or correlation functions on the imaginary-time axis or Matsubara frequencies, which then need to be analytically continued to the real axis to extract dynamical information about the system of interest.
	One important example of analytic continuation is obtaining the spectral function~$A(\omega)$ at real frequencies from  finite-temperature Green function values $\mathcal{G}(\tau)$  at  imaginary times~$\tau\in [0,\beta]$, where the two functions are related by
	\begin{equation}\label{eq:tau_spectral}
		\mathcal{G}(\tau) = -\frac{1}{2\pi}\int d\omega \; \frac{\mathrm{e}^{-\tau\omega}}{ 1 + \, \mathrm{e}^{-\beta \omega} } \; A(\omega), 
	\end{equation}
	and $\beta=1/T$ is the inverse temperature. 
	Another example is determining the optical conductivity spectrum $\sigma(\omega)$ from the current-current correlation function $\Pi(i\omega_m)$  evaluated at bosonic Matsubara frequencies $\omega_n=2m\pi/\beta$.
	The relation between the two reads
	\begin{equation}\label{eq:matsubara_conductivity}
		\Pi(i\omega_m) = \frac{2}{\pi} \int_0^{\infty} d\omega \;  \frac{\omega^2}{\omega_m^2+\omega^2} \ \sigma(\omega)\;.
	\end{equation}
	
	In general, the analytic continuation problem reduces to solving an integral equation.
	The difficulty is that, in the presence of noise, this is an inherently ill-posed problem. 
	When evaluating the data on the imaginary axis, oscillations and sharp features in the spectrum get smoothed and noise gets damped due to the integration.
	This makes the inverse problem of reconstructing the details of the spectrum extremely challenging. 
	Without regularization,  small noise on the data leads to catastrophically large errors on the best data-fitting spectrum.
	
	There are different approaches to tackle this problem including the maximum entropy method~\cite{Silver90, Jarrell96, Gunnarsson10, Jarrell12, Tremblay15}, the average spectrum method~\cite{White91, Sandvik98, Beach04, Gunnarsson07, Syljuasen08, Fuchs10, Sandvik16, Ghanem20}, Pade approximation techniques~\cite{Vidberg77, Beach00, Ostlin12, Schot16}, and some recent machine-learning based approaches~\cite{Arsenault14, Arsenault16,Yoon18, Fournier20}.
	The most commonly used approach is the maximum entropy method (MaxEnt), which is rooted in Bayesian inference. 
	It tries to find a spectrum by balancing the fit to the data and the entropy relative to some default model.
	This entropy term acts as a regularization that penalizes deviations from the featureless default model and thus suppresses rapid oscillations that otherwise would dominate the solution.
	
	The average spectrum method (ASM) is an alternative Bayesian approach with the following premise:
	Since the data is not exact, all spectra that fit the data equally-well, up to the noise level on that data, should be considered equally.
	As a result, ASM integrates over all admissible spectra weighted by how well they fit the data. 
	The average spectrum method makes no assumptions about the smoothness of the spectrum and any regularization comes from averaging only, which is expected to smooth out details not supported by the data: larger noise leads to more smoothing. 
	
	We continue here our work published in Ref.~\cite{Ghanem20}, which in the following will be referred to as ASM1.
	In ASM1, we showed that a naive discretization of the functional integral involved in the average spectrum method does not produce unique results. The results are biased by the discretization grid on which the spectrum is represented.
	We constructed the grids explicitly by mapping the real-frequency axis to the unit interval using a density function and then discretizing this interval using a regular grid.
	We found that the density function of the grid points plays the role of a default model while the number of grid points acts as a regularization parameter.
	We proposed a practical recipe for choosing a reliable grid by comparing the results of different grid densities and choosing the one with the least dependence on the number of grid points.
	
	In the present paper, we generalize the average spectrum method by releasing the grid points and sampling their position from a prior grid density.
	Although, we still need to specify a grid density and a number of points, we show that this the bias is significantly reduced and we observe that it depends mainly on the width rather than the shape of its density.
	The proper width can be chosen according to the same type of recipe used earlier for the fixed grid.
	We can, however, go further and extend the method to sample over a whole class of grid densities of variable widths.
	The method is then able to automatically relocate the grid points and concentrate them into the important region of the real-frequency axis. 
	Test cases show that this width-sampling method gives good results resolving the features of the spectrum without the need for fine-tuning the grid.
	
	\section{Average Spectrum Method}
	\subsection{Background}
	Mathematically, the analytic continuation problem can be formulated as a Fredholm integral equation of the first kind
	\begin{equation}
		g(y) = \int dx\ K(y, x) f(x)\;,
	\end{equation}
	where the left-hand side $g(y)$ represents QMC data, while the integral kernel $K(y, x)$ is a continuous function known analytically.
	The goal is estimating  the spectrum $f(x)$, an integrable non-negative function.
	
	QMC provides noisy and incomplete samples of the data evaluated at a finite number of $y$-coordinates.
	Using the central limit theorem, one can assume that the exact data vector $\vec{g}$ has a Gaussian distribution with mean equal to the sample mean $\vec{\bar{g}}$ and covariance matrix equal to the sample covariance matrix $\vec{C}$.
	Then the likelihood of a spectrum $f$ being the exact one is proportional to
	\begin{equation}
		\exp\left(-\frac{1}{2} \big(\vec{\bar{g}}-\vec{g}[f] \big)^{\dagger} \vec{C}^{-1}  \big(\vec{\bar{g}}-\vec{g}[f] \big)\!\right)  \eqqcolon  \mathrm{e}^{-\chi^2[f]/2},
	\end{equation}
	where $\vec{g}[f]$ is the data corresponding to the spectrum $f$ and $\chi^2[f]$ is its fit to the measured data.
	Maximizing the likelihood (or equivalently minimizing the fit $\chi^2$) gives a spectrum dominated by diverging amplitudes that are extremely sensitive to the noise of the data.
	This ill-posedness constitutes the primary difficulty of the analytic continuation problem.
	
	The average spectrum method uses the likelihood alongside our prior knowledge about the non-negativity of the spectrum and computes a weighted average over all non-negative spectra as an estimate of the true spectrum
	\begin{equation}\label{eq:asm}
		f_\text{ASM}(x) \propto \int\limits_{f(x)\geq0} \mathcal{D}f \; \mathrm{e}^{-\chi^2[f]/2} f(x) \;.
	\end{equation}
	Other exact prior knowledge, like sum rules, can be easily incorporated by restricting the averaging to the spectra satisfying them.
	
	Despite the apparent elegance of this functional integral formulation, we found in ASM1 that it is not a well-defined expression because the result depends on how the spectrum $f(x)$ is discretized. 
	In ASM1 the discretization was specified by a grid density function $\rho(x)$ and the total number of grid points $N$.
	Using $\rho(x)$, we mapped the domain of the variable $x$ into a unit interval of an auxiliary variable $z(x) \coloneqq \int_{x_\text{min}}^{x} dx^\prime \rho(x^\prime)$, discretized it uniformly with $N$ points, and mapped the points back to $x$.
	We showed for such grids that the density $\rho(x)$ plays the role of a default mode, while $N$ plays the role of a regularization parameter.
	
	To demonstrate this role of the grid density $\rho(x)$ and simultaneously introduce some notation, we repeat the  argument of appealing to symmetry when the data contains no information other than normalization to unity.
	In that case, the spectral integrals over different grid intervals $\mathcal{I}_i=[\bar{x}_i,\bar{x}_{i+1}]$ should be equal
	\begin{equation}\label{eq:f_int}
		\bar{f}_i \coloneqq \int_{\bar{x}_i}^{\bar{x}_{i{+}1}}dx\, f(x) = \frac{1}{N}\;.
	\end{equation}
	The widths of these intervals are inversely proportional to the grid density
	\begin{equation}
		\Delta x_i \approx \frac{1}{N\rho(x_i)}\;,
	\end{equation}
	with $x_i\in\mathcal{I}_i$ being the grid point representing the $i$-th interval.
	Therefore, in the absence of data, the estimated mean values of the spectrum are equal to the grid density
	\begin{equation}
		f(x_i)  = f_i  \coloneqq \frac{ \bar{f}_i}{\Delta x_i}  \approx  \rho(x_i)\;,
	\end{equation}
	justifying calling it a default model.
	
	\subsection{Formalism}
	To spell out the dependence of the average spectrum method on the discretization grid explicitly, let us parameterize the spectrum by its grid points $\vec{x}$ and its integrals over the grid intervals $\vec{\bar{f}}$, defined in \eqref{eq:f_int}.
	Together they are enough to determine the data (up to a discretization error) without knowing the  details of the spectrum inside the grid intervals.
	This follows from the first mean-value theorem for integrals, which states that for any non-negative integrable function $f(x)$, there is a specific point $x_i^\star \in [\bar{x}_i, \bar{x}_{i+1}]$ such that
	\begin{equation}
		\int_{\bar{x}_i}^{\bar{x}_{i+1}} dx\ K(y, x) f(x) =  K(y, x_i^\star) \int_{\bar{x}_i}^{\bar{x}_{i+1}} dx\ f(x)\;.
	\end{equation}
	Using this, we can approximate the data as following
	\begin{equation} \label{eq:data}
		\vec{g}[\vec{\bar{f}}, \vec{x}] \approx \sum_i K(\vec{y};x_i) \bar{f}_i \eqqcolon  \vec{K}[\vec{x}]\ \vec{\bar{f}} \;.
	\end{equation}
	The approximation comes from using the grid points $x_i$ instead of the unknown optimal points $x_i^\star$, which depend on both the spectrum $f(x)$ and the kernel $K(x,y)$.
	The approximation error is proportional to the difference between the maximum and minimum values of the kernel  $K(x, y)$ inside each interval.
	Since the kernel is a continuous function of $x$, this error gets smaller, the smaller the intervals are and the smoother the kernel is.
	Using a fine enough grid, the error becomes so small that it is negligible in comparison to the noise on the data.
	
	Contrary to the data, the spectral integrals $\vec{\bar{f}}$ do not provide sufficient information to determine the spectrum completely and some assumptions about its behavior inside the grid intervals are needed.
	We may also need to specify how to go from grid points $\vec{x}$ to the intervals edges $\vec{\bar{x}}$.
	To stay as general as possible, we encode whatever assumptions we have in the object $f(\vec{\bar{f}}, \vec{x}; x)$, which maps a set of gird points  $\vec{x}$ and integrals $\vec{\bar{f}}$ into a non-negative integrable function of the continuous variable $x$.	
	A simple choice is using delta functions located at the grid points
	\begin{equation}\label{eq:delta_rep}
		f(\vec{x},\vec{\bar{f}};x) = \sum_i \bar{f}_i\,\delta(x-x_i)\;
	\end{equation}
	another is a piece-wise constant function
	\begin{equation}\label{eq:const_rep}
		f(\vec{x},\vec{\bar{f}};x) = \sum_i \frac{\bar{f_i}}{\bar{x}_{i+1}{-}\bar{x}_i}\Big(\Theta(x{-}\bar{x}_i)-\Theta(x{-}\bar{x}_{i+1})\Big).
	\end{equation}
	The advantage of the above general formulation is that we do not need to know the exact form of $f(\vec{\bar{f}}, \vec{x}; x)$  when sampling the spectral functions, it becomes relevant only when averaging them, which will be discussed in Sec.~\ref{sec:binning}.
	
	The functional integral of the average spectrum can now be written as a multidimensional integral over parameterized functions $f(\vec{\bar{f}},\vec{x}; x)$ given the mean data vector $\vec{\bar{g}}$, the covariance matrix $\vec{C}$ and the gird points $\vec{x}$.
	The averaging is carried over all spectral integrals $\vec{\bar{f}}$ on that grid, weighted by their fit 
	\begin{equation}\label{eq:asm_discrete}
		f_\text{ASM}(\vec{x}; x) \propto \int\limits_{0}^{\infty}  d\vec{\bar{f}} \; \mathrm{e}^{-\chi^2[\vec{\bar{f}}, \vec{x}]/2}\ f(\vec{\bar{f}},\vec{x}; x)\;,
	\end{equation}
	where the dependence of the fit $\chi^2$  on the grid points  $\vec{x}$ is through the discretized kernel matrix $\vec{K}$ evaluated at these points 
	\begin{equation}
		\chi^2[\vec{\bar{f}}, \vec{x}] \coloneqq \left(\vec{\bar{g}}-\vec{K}[\vec{x}]\ \vec{\bar{f}}\right)^{\dagger} \vec{C}^{-1}  \left(\vec{\bar{g}}-\vec{K}[\vec{x}]\ \vec{\bar{f}} \right)\;.
	\end{equation}
	
	\section{Releasing Grid Points}
	The form of Eq.~\eqref{eq:asm_discrete} spells out the grid dependence of ASM explicitly and is suggestive of a straightforward extension of the method. 
	Instead of having the grid points fixed at regular intervals based on some density function $\rho(x)$, let us sample them freely from this distribution and average the results
	\begin{equation}\label{eq:gstochs}
		f_\text{ASM}(\rho, N; x) \!\propto 
		\!\!\int\!\! d\vec{x} \prod_{i=1}^{N} \rho(x_i) \!\!\int\limits_{0}^{\infty}\!  d\vec{\bar{f}} \, \mathrm{e}^{-\frac12\chi^2[\vec{\bar{f}}, \vec{x}]} f(\vec{\bar{f}},\vec{x}; x).
	\end{equation}
	Although we still need to specify the density function $\rho(x)$ and the number of grid points $N$ as before, we expect the effect on the results  to be weaker than in the fixed-grid scenario because  the data is now allowed to influence the positions of the grid points during the sampling.
	
	It is worth noting that sampling grid points has been done before in the context of the average spectrum method in Refs.~\cite{Beach04, Fuchs10}. 
	In these papers, the spectrum was represented as a superposition of delta functions whose both weights and \emph{positions} are sampled.
	However, it was implicitly assumed that sampling the positions is a technical detail and that the result would be the same as the typical average spectrum method with fixed positions~\cite{White91, Sandvik98, Syljuasen08}.
	This is probably due to the mistaken belief in the existence and uniqueness of the functional integral of Eq.~\eqref{eq:asm}. 
	As shown in ASM1, this functional integral does not exist and the result depends on the discretization grid.
	In Sec.~\ref{sec:density_dependece} we will show that the results, in fact, depend on whether the grid points are sampled and rather improves significantly.
	But before that, we will describe an efficient algorithm for performing the sampling and how the spectra of different grids are averaged.
	
	\subsection{Sampling algorithm}\label{sec:sampling}
	The multidimensional integral in Eq.~\eqref{eq:gstochs} is evaluated using a Monte Carlo sampling algorithm.
	We start from some initial spectrum on an initial grid. In practice, we use the same grid as in the fixed-grid case, i.e., we choose the grid points at regular intervals based on the density $\rho(x)$ as in ASM1.
	We also use the non-negative least squares (NNLS) solution on that grid as the initial spectrum~\cite{Lawson95}.
	
	The spectral integrals are then updated on the current grid using the blocked-mode sampling introduced in ASM1.
	Given the spectral integrals, the grid points are updated one at a time using the Metropolis-Hastings algorithm explained below.
	All samples of $\vec{\bar{f}}$ and $\vec{x}$ are stored during the sampling, and the average spectrum $f_\text{ASM}(\rho, N; x)$ can be evaluated later at any point $x$ by  evaluating each sampled model $f(\vec{\bar{f}}, \vec{x}; x)$ at that point and averaging the results. 
	
	The Metropolis-Hastings algorithm for sampling grid points has the following acceptance ratio
	\begin{equation}\label{eq:acc_ratio}
		r = \frac{\mathrm{e}^{-{\chi}^2(x_i^\prime)/2} }{\mathrm{e}^{-{\chi}^2(x_i)/2} } \frac{\rho(x_i^\prime)}{\rho(x_i)}\frac{q(x_i^\prime\to x_i)}{q(x_i\to x_i^\prime)}\;,
	\end{equation}
	where $q(x_i\to x_i^\prime)$ is the proposal distribution of moving  grid point $i$ from an old position $x_i$ to a new position $x^\prime_i$.
	As a proposal distribution, we use a Gaussian approximation of the data factor  $\mathrm{e}^{-\chi^2/2}$.
	It is derived by writing the data fit as a function of  $x_i^\prime$
	and expanding the kernel around $x_i$.
	Keeping only terms to second-order in $x^\prime_i$, we obtain
	\begin{equation}
		\begin{split}
			\chi^2(x_i^\prime) \approx\quad & \vec{r}^\mathrm{T}\vec{r} -2 \left[\vec{r}^\mathrm{T}\partial\vec{K}_i \bar{f}_i\right] \left[x^\prime_i - x_i\right]+ \\
			& \left[\bar{f}_i^2 \partial\vec{K}_i^\mathrm{T}\partial\vec{K}_i - \bar{f}_i \vec{r}^\mathrm{T} \partial^2\vec{K}_i \right] \left[x^\prime_i - x_i\right]^2\;,
		\end{split}
	\end{equation}
	where $\vec{r} \coloneqq \vec{g} - \sum_j \vec{K}(x_j)   \bar{f}_j $ is the old residual vector and $\partial\vec{K}_i \coloneqq \partial_x\vec{K}(x_i),\ \partial^2\vec{K}_i \coloneqq \partial_x^2\vec{K}(x_i)$ are kernel derivatives.
	By completing the squares, the data fit can be written in the following suggestive form
	\begin{align}\label{chiApprox}
		\chi^2(x_i^\prime) &\approx (x_i^\prime - \mu_\chi)^2/\sigma_\chi^2 + \text{const.} 
	\end{align}
	where $ \mu_\chi \coloneqq x_i + \sigma^{2}_\chi \bar{f}_i \vec{r}^\mathrm{T} \partial\vec{K}_i$ is the mean of the Gaussian approximation and $\sigma_\chi^{-2} \coloneqq  \bar{f}_i^2 \partial\vec{K}_i^\mathrm{T}\partial\vec{K}_i - \bar{f}_i \vec{r}^\mathrm{T} \partial^2\vec{K}_i$ is its width.
	Using this as a proposal probability gives in general a high acceptance rate because the data is typically the dominating factor in the  acceptance ratio~\eqref{eq:acc_ratio}.
	However, when a grid point is far away from zero or its weight is very small, the Gaussian \eqref{chiApprox} is quite wide and the prior density $\rho(x)$ becomes important.
	To account for such cases, we combine the data Gaussian with another one, centered at the old position $x_i$, whose width we choose equal to the width of the prior density.
	The latter is used by itself in the proposal probability should the data fit have negative curvature so that the approximation \eqref{chiApprox} fails.
	
	Samples are produced by iteratively sampling all the grid points  $\vec{x}$ followed by sampling all the spectral integrals $\vec{\bar{f}}$.
	The grid points are sampled one after the other  in  a random order  using the above algorithm.
	The spectral integrals are sampled using blocked-mode sampling with a random block size.
	The movement of grid points implies that the kernel matrix is changing, so the singular value decomposition (SVD) of its blocks should be recalculated in each iteration after all the grid points have been updated.
	This decomposition costs $\mathcal{O}(N_b^2M)$ where $N_b$ is the block size and $M$ is the data vector size.
	Since this is computationally expensive, we restrict the block size to a maximum value.  
	For a fixed grid, this leads to more correlation between samples because the global updates using larger blocks are skipped.
	For a released grid, however, the effect is not as severe because the movement of the grid points compensates for the lack of these global updates.
	We found that a maximum block size of $32$ provides a good balance between the cost of each sample and the correlation between samples.
	Due to grid points sampling and SVD, the computational cost of a single sample of the average spectrum method using released grid points is larger than that of the fixed grid.
	However, we found that the samples of the released-grid calculations are much less correlated than those on a fixed-grid so that the total cost is effectively similar.
	The execution time for any of the results presented later in this paper does not exceed 2 minutes on a typical modern laptop.
	
	\subsection{Binning and averaging}\label{sec:binning}
	Averaging samples requires evaluating the sampled spectra on some fixed grid.
	We call this grid the binning grid to distinguish it from the sampled one $\vec{x}$.
	Let us denote its intervals, the bins, as $\mathcal{B}_i$.
	The binning would  be different depending on the mapping $f(\vec{\bar{f}}, \vec{x}; x)$.
	Assuming the delta functions representation of Eq.~\eqref{eq:delta_rep}, the $i$-th bin average is computed as
	\begin{equation}\label{eq:delta_binning}
		f_i \approx \frac{1}{L}  \sum_{k=1}^L   \frac{1}{\operatorname{len}(\mathcal{B}_i)} \sum_{x^k_j \in \mathcal{B}_i}  \bar{f}^k_j\;,
	\end{equation}
	where the superscript $k$ is indexing the samples and $L$ is the total number of samples.
	
	Alternatively, we can assume a constant value inside each interval $\mathcal{I}_j^k$ of the $k$-th grid sample, as done in Eq.~\eqref{eq:const_rep}.
	This implies that the corresponding spectral integral $f^k_j$ should be split proportionally among the bins that intersects this interval
	\begin{equation}
		f_i  \approx \frac{1}{L}  \sum_{k=1}^L  \frac{1}{\operatorname{len}(\mathcal{B}_i)} \sum_{j=1}^N \frac{\operatorname{len}(\mathcal{B}_i\cap\mathcal{I}^k_j)}{\operatorname{len}(\mathcal{I}^k_j)}\  \bar{f}^k_j\;.
	\end{equation}
	This type of binning can be thought of as a linear interpolation of \eqref{eq:delta_binning} and thus leads to an average with less noise from binning.
	Nevertheless, whatever binning we use, the averages are similar when using reasonably large grid sizes $N$.
	For simplicity, we typically use the delta binning.
	
	Note that the bin size also affects the statistical error of its average.
	Larger bins contain more sampled grid points and thus have lower fluctuations and less noisy averages.
	Therefore, in practice we use the following binning, which gives roughly equal error bars across the bins: aggregate all the grid samples and choose the bins such that each bin contains roughly the same number of grid points. 
	When we want to compare with fixed-grid ASM, we use, however, the fixed grid for binning, making comparisons easier.
	Also note that the error-bars of a binning grid will not change when doubling the number of grid points and halving the number of samples.
	
	\section{Density Dependence}
	\label{sec:density_dependece}
	
	\begin{figure}
		\center
		\includegraphics[width=0.95\columnwidth]{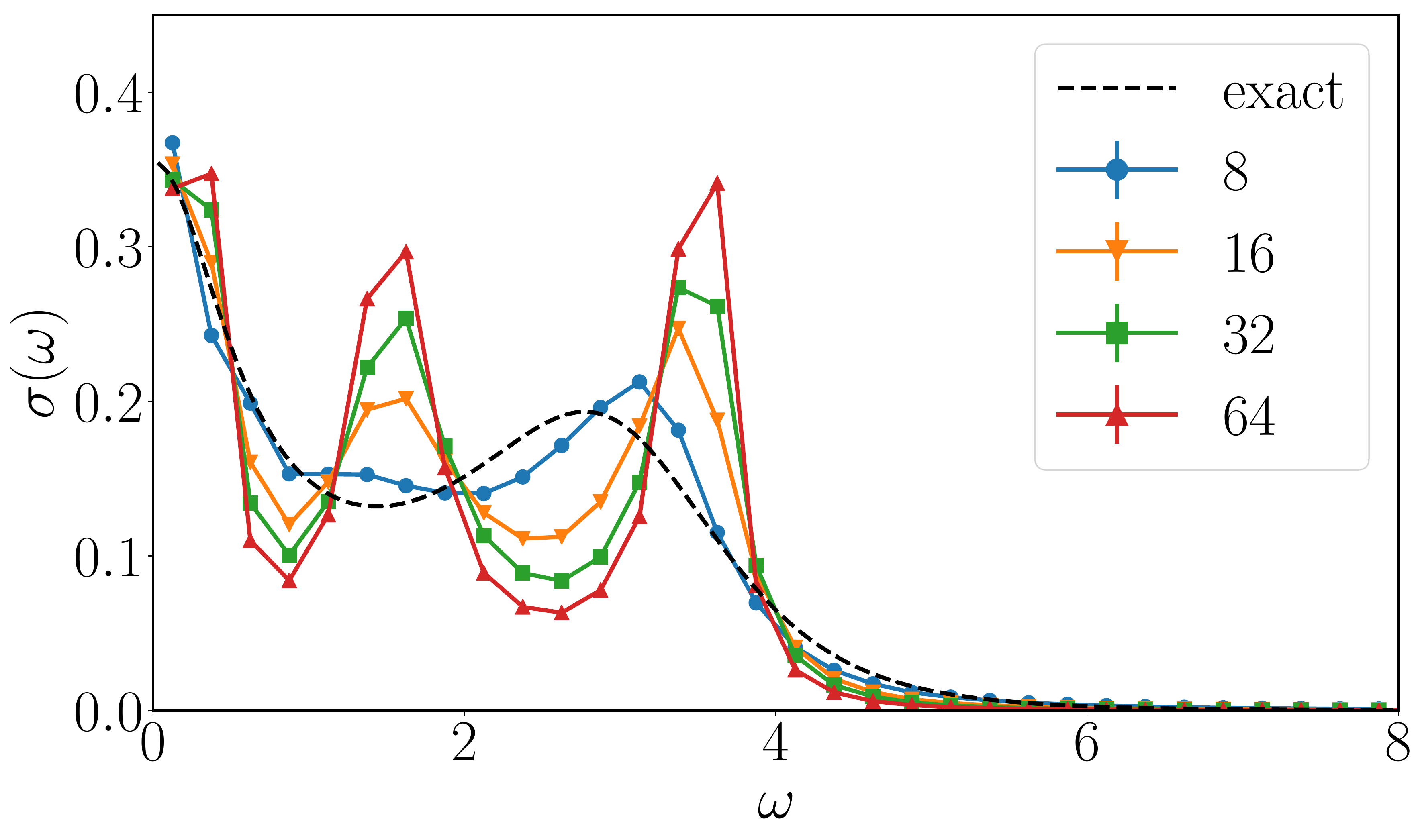}
		\includegraphics[width=0.95\columnwidth]{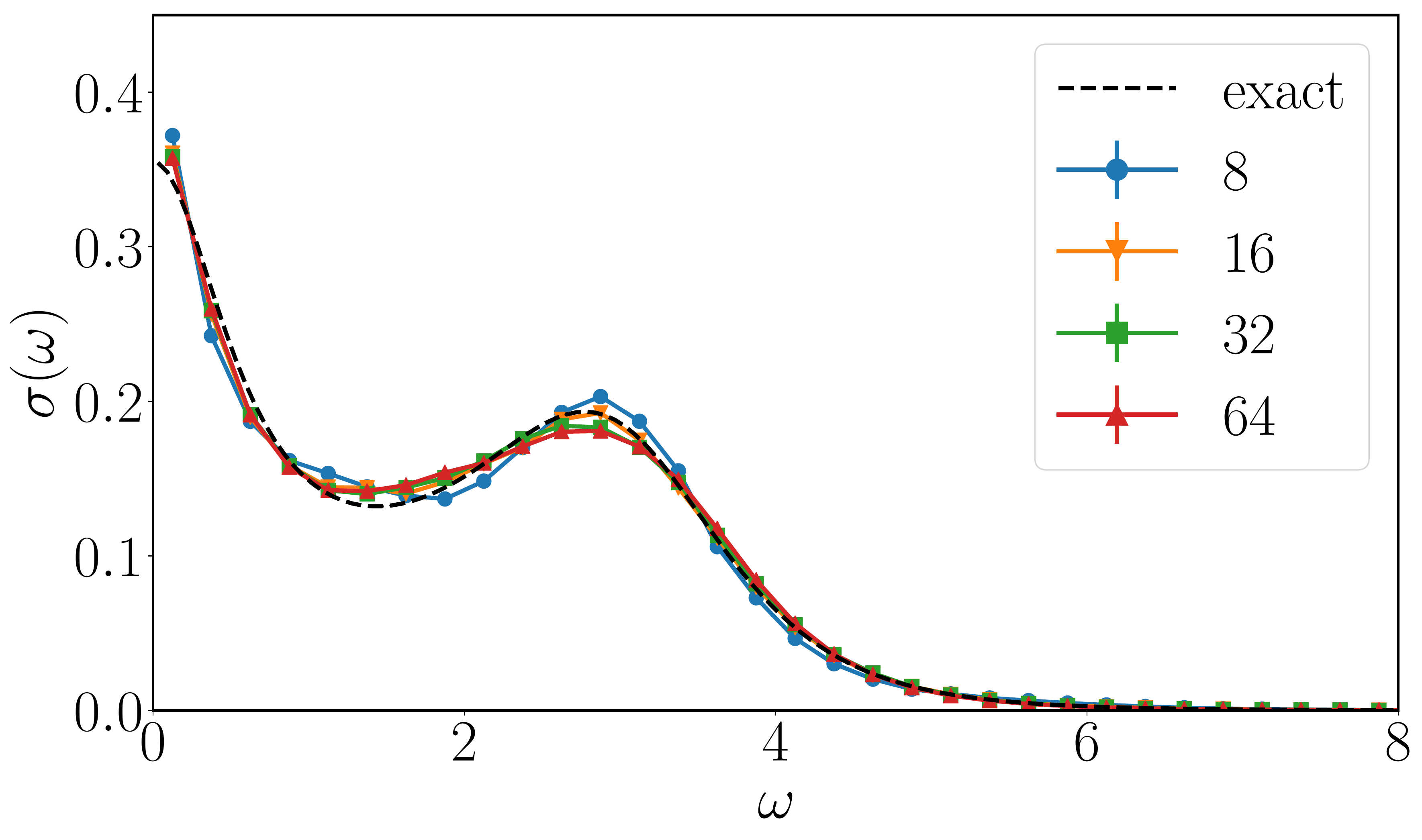}
		\caption{\label{fig:cutoff}
			Optical conductivity $\sigma(\omega)$ obtained using fixed-grid ASM (top) and released-grid ASM (bottom).
			Uniform grid densities with increasing cutoff $\omega_\text{max}$ (label) are used.
			The number of grid points is proportional to the cutoff: $N = 4\ \omega_\text{max}$.
			For ease of comparison, the samples of the released calculations are binned and averaged on the grids of the corresponding fixed calculations.
		}
	\end{figure}
	
	\begin{figure}
		\center
		\includegraphics[width=0.95\columnwidth]{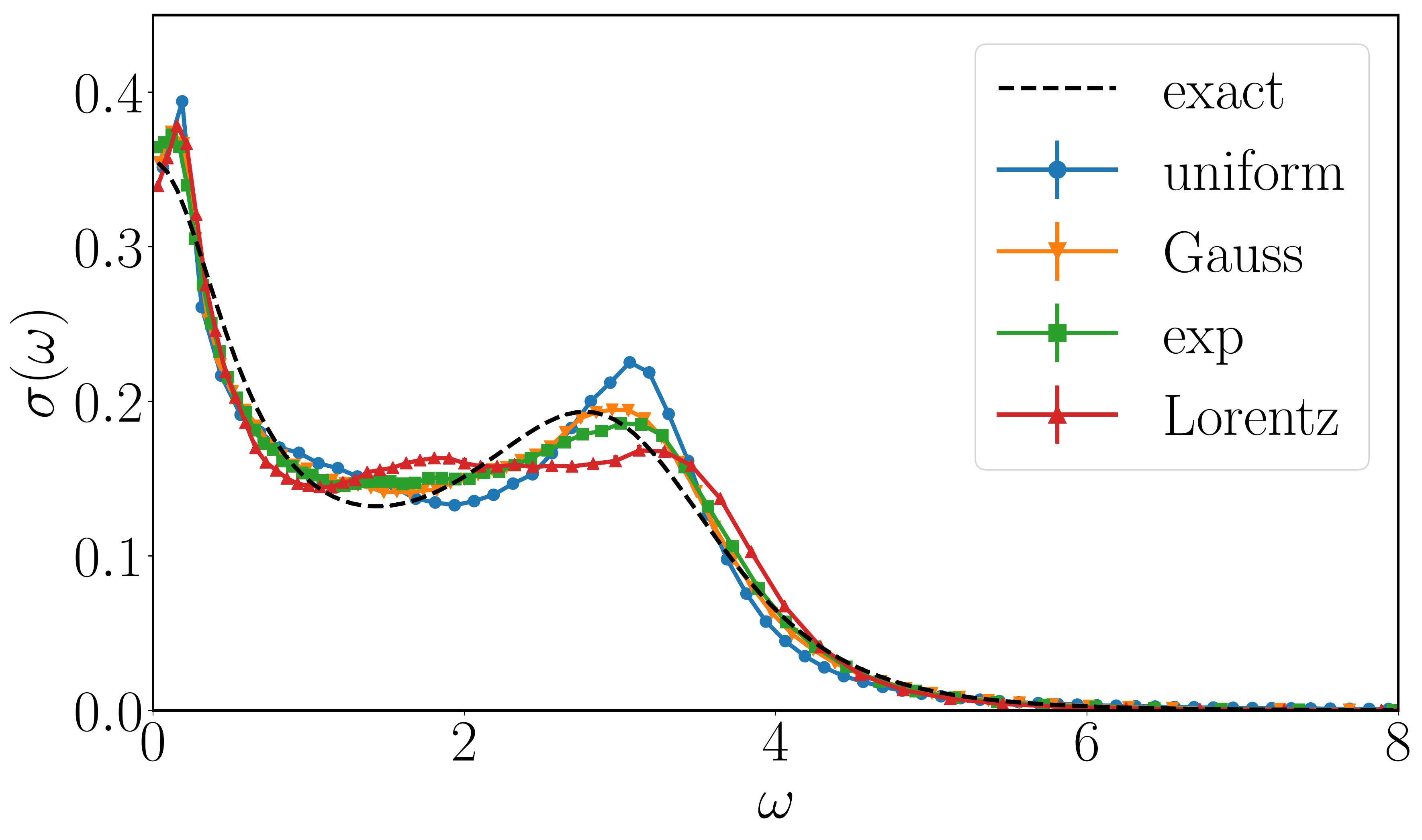}
		\includegraphics[width=0.95\columnwidth]{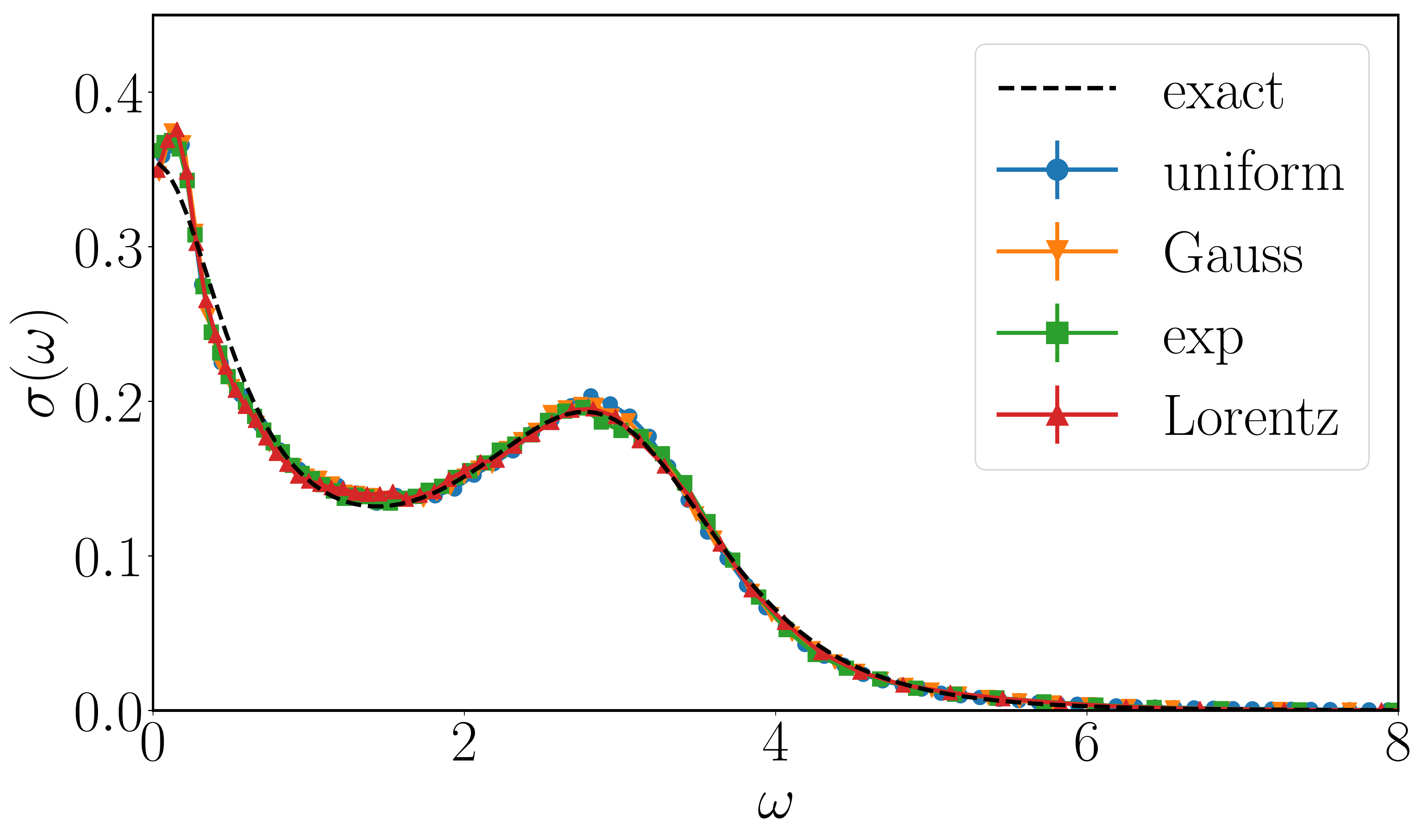}
		\caption{\label{fig:density} 
			Optical conductivity $\sigma(\omega)$ obtained using fixed-grid ASM (top) and released-grid ASM (bottom).
			The following grid densities (labels) are used: uniform ($\omega_\text{max} = 8$), Gaussian ($\alpha=4$), exponential ( $\beta=3$) and a Lorentzian ($\gamma=2.5$).
			The number of grid points is $N=64$.
			For ease of comparison, the samples of the released calculations are binned and averaged on the grids of the corresponding fixed calculations.
		}
	\end{figure}
	
	To study the effect of releasing the grid points, we choose a test case first introduced in Ref.~\cite{Gunnarsson10b}, which we studied further using fixed-grids in ASM1.
	We want to reconstruct an optical conductivity given by
	\begin{equation}\label{model}
		\sigma(\omega) = \frac{1}{1{+}(\omega/\Gamma_e)^6}\!\! \sum_{p=0,\pm1} \frac{W_{|p|}}{1+((\omega{+}\mathrm{sgn}(p)\varepsilon_{|p|})/\Gamma_{|p|})^2} 
	\end{equation}
	where the sum has three terms: a peak of weight $W_0=0.3$ and width $\Gamma_0=0.6$ centered at zero, and two peaks of weight $W_1=0.2$ and width $\Gamma_1=1.2$ centered at $\omega=\pm\varepsilon_1=\pm 3$. 
	All terms are multiplied by a damping factor with $\Gamma_e=4$ for a faster decay at large frequencies.
	To get the necessary data for analytic continuation, we compute  the imaginary-frequency correlation function $\Pi(i\omega_m)$ analytically using Eq.~\eqref{eq:matsubara_conductivity} on the first $60$ Matsubara frequencies  $\omega_m=2m\pi/\beta$ with inverse temperature $\beta=15$.
	We then add relative Gaussian noise with a standard deviation of $10^{-3}$ to simulate the noise in real QMC data.
	
	In Fig.~\ref{fig:cutoff}, we compare ASM results using fixed and released grid points.
	The same grid densities and numbers of grid points were used in both cases: 
	uniform grid densities with increasing cutoffs: 8, 16, 32, and 64 and correspondingly increasing number of points $N= $ 32, 64, 128,  and 256.
	It is clear that using a fixed uniform grid biases the results significantly, leading to more pronounced spurious features as the cutoff increases. 
	This completely disappears when the grid points are released so that the cutoff has negligible effect on the result.
	In Fig.~\ref{fig:density},  we also show results for different grid densities with comparable widths.
	Also here, by releasing the grid points, the influence of the shape of the grid density is reduced significantly.
	This is understandable as now the grid points can move to the region where the spectrum is concentrated, allowing for the data to override the prior information encoded in the grid density.
	
	From this we might conclude that sampling the grid points solves the bias problem. We know, however, that in the absence of data except for a sum rule on the spectrum, averaging spectra on any grid just gives a result proportional to the grid density function $\rho(x)$. Thus, also released-grid ASM must have a default model bias.
	
	\begin{figure}
		\center
		\includegraphics[width=0.95\columnwidth]{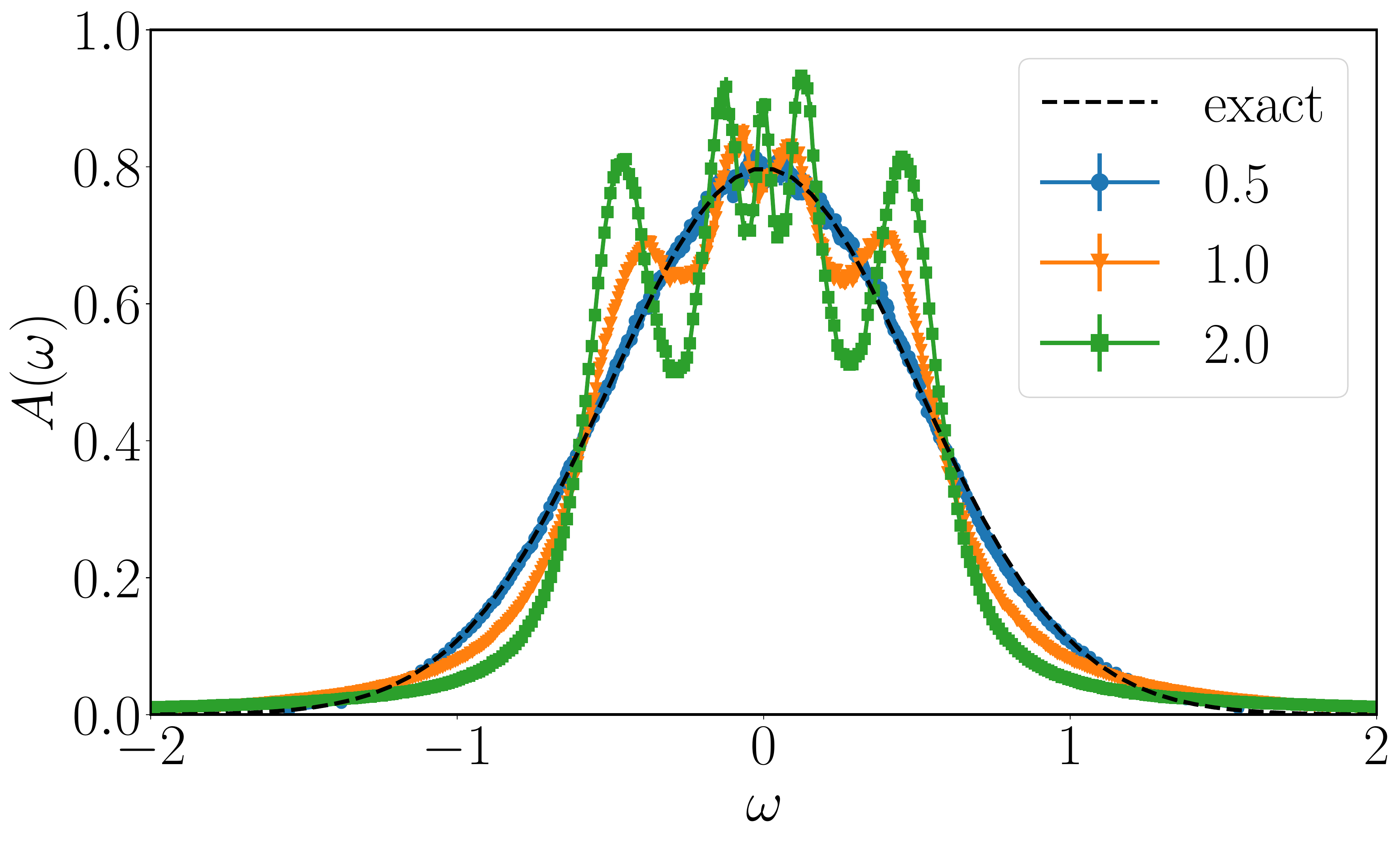}
		\includegraphics[width=0.95\columnwidth]{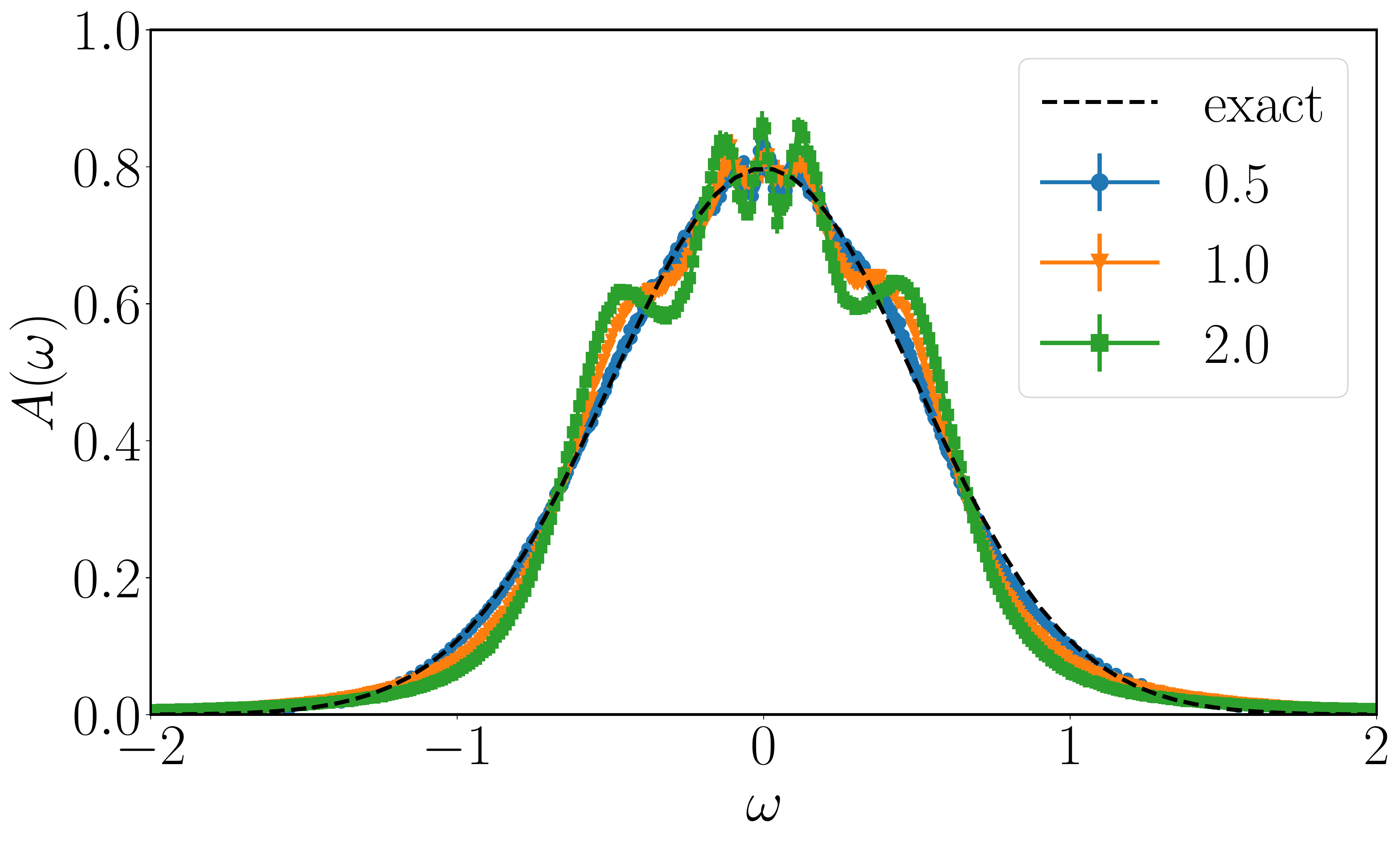}
		
		\caption{\label{fig:width}
			Spectral function $A(\omega)$ obtained using fixed-grid ASM (top) and released-grid ASM (bottom). Gaussian grid densities with increasing width (label) are used.
			The number of grid points is $N=512$.
			For ease of comparison, the samples of the released calculations are binned and averaged on the grids of the corresponding fixed calculations.
		}		
	\end{figure}
	
	\begin{figure}
		\center
		\includegraphics[width=0.95\columnwidth]{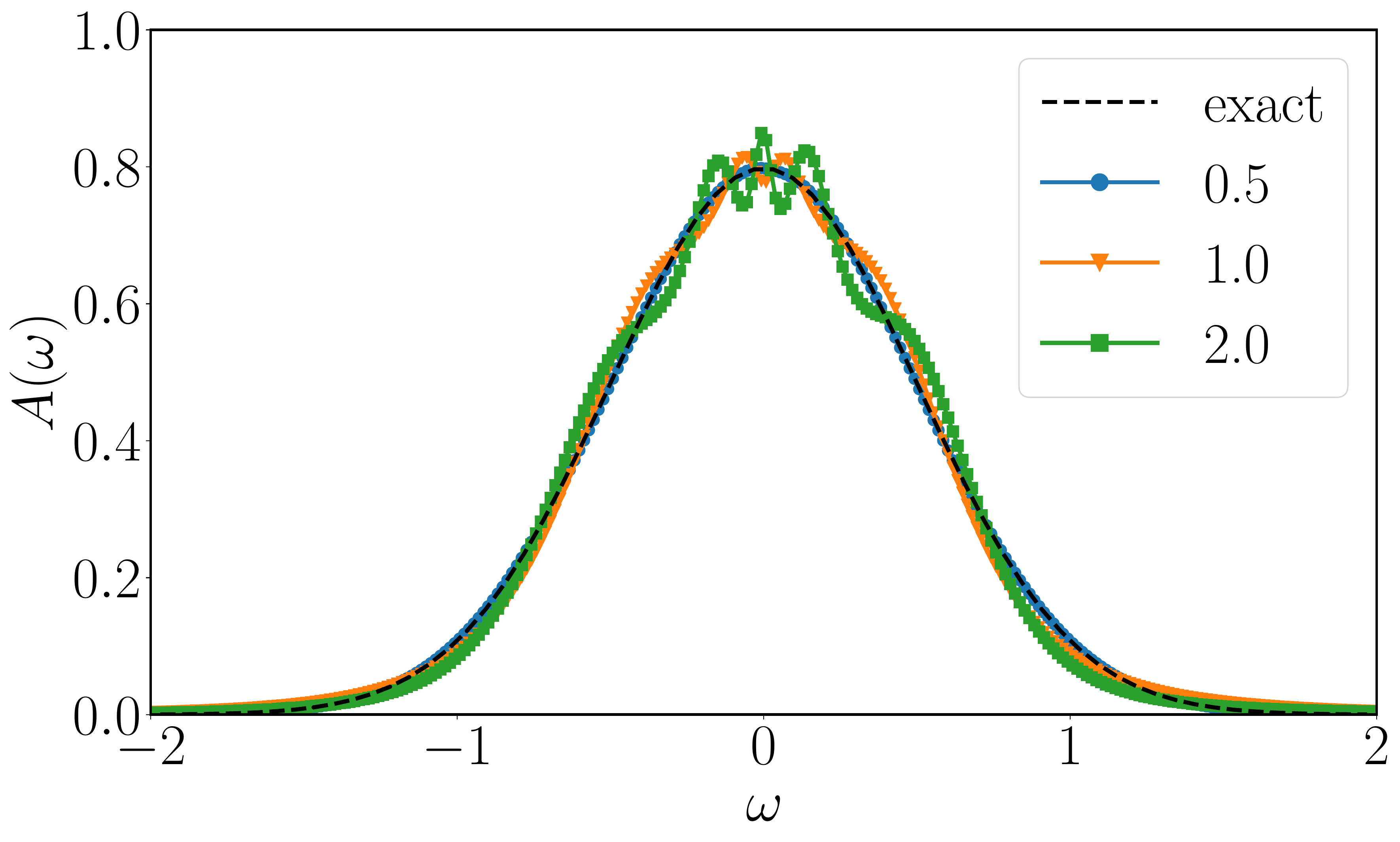}
		\caption{\label{fig:maxent}%
			Spectral function $A(\omega)$ obtained using MaxEnt with Gaussian default models used in Fig.~\ref{fig:width}.
			The results are obtained using Bryan's method implemented in Ref.~\cite{Levy17}.
			Other methods for choosing the regularization parameter, i.e., classic and historic MaxEnt give indistinguishable results. 
		}
	\end{figure}
	
	To see this bias clearly, we consider a somewhat contrived test case:  we seek to recover a Gaussian spectral function of width $0.5$ from Green function data computed using Eq.~\eqref{eq:tau_spectral}. 
	The data are evaluated on 60 $\tau$-points equally spaced in the interval $[0, \beta]$ with $\beta=50$.
	As before, we add relative Gaussian noise with a standard deviation of $10^{-3}$.
	One might think that reconstructing such a featureless Gaussian should be a trivial and boring task.
	However, it turns out that avoiding spurious features in this setting is more challenging than anticipated.
	In Fig.~\ref{fig:width}, we show ASM results using fixed and released grids.
	We use a Gaussian grid density with different widths: 0.5, 1.0, 2.0.
	When the default model (grid density) width equals the width of the exact spectrum $0.5$,  both methods give perfect results as expected.
	As the width increases, spurious features start to develop quickly.
	Although these features are milder for  released-grid than for fixed-grid ASM, they are still clearly visible, in contrast with the earlier optical conductivity case.
	This indicates that the data here is weaker in forcing the grid points to stay in the frequency region of interest.
	A practical solution for choosing the best width is to use the same criterion as the one employed in choosing the fixed-grid in ASM1:
	we choose the grid density with the best fit to the data, which in this case would single out the width $0.5$.
	Still, this approach requires considering a number of grid densities $\rho(x)$ of different widths and choosing the best one.
	In the next section, we describe a method that avoids choosing a specific width altogether.
	It is worth mentioning that the bias caused by the default model is not unique to ASM. 
	Also MaxEnt, which is known for its smooth results, produces in this case spurious features when the width of the default model is larger than it is supposed to be.
	They are, however, somewhat milder than those of the two flavors of ASM (see Fig.~\ref{fig:maxent}).
	
	\section{Averaging Width of Grid Density }\label{sec:width_avg}
	Instead of finding the width giving the best fit to the data, we propose here an alternative that is more in the spirit of the average spectrum method: Instead of choosing some width {\it a priori}, 
	we average over the width parameter of the grid density.
	Integrating over this parameter in Eq.~\eqref{eq:gstochs} requires evaluating the  integral
	\begin{multline}
		\int\! dw\!   \int\!   d\vec{x}  \prod_{i=1}^{N} \rho(w; x_i) \!\int\limits_{0}^{\infty}\!  d\vec{\bar{f}} \, \mathrm{e}^{-\chi^2[\vec{\bar{f}}, \vec{x}]/2}\, f(\vec{\bar{f}},\vec{x}; x) .
	\end{multline}
	One way could be sampling the width directly using Metropolis-Hasting, but that would be inefficient: updating the width would change the prior probabilities for all the grid points.
	So for a large number of grid points, one would be forced to take very small updates of the width to achieve a reasonable acceptance rate.
	The more grid points, the less efficient the sampling is. 
	There is, however, a much more elegant and efficient solution.
	
	We notice that unlike the grid points $\vec{x}$ and spectral integrals $\vec{\bar{f}}$, the width parameter $w$ is not directly related to the data.
	Therefore, the above expression can be rearranged such that the integral over the width is a function of the grid points only
	\begin{equation}  \label{eq:width_avg}
		\int\!  d\vec{x}  \underbrace{\int\! dw  \prod_{i=1}^{N} \rho(w; x_i)}_{=:P(\vec{x})}  \int\limits_{0}^{\infty}\!  d\vec{\bar{f}} \, \mathrm{e}^{-\chi^2[\vec{\bar{f}}, \vec{x}]/2}\, f(\vec{\bar{f}},\vec{x}; x).
	\end{equation}
	We can then perform the width integral $P(\vec{x})$ analytically, e.g. for the family of density functions \begin{equation}
		\rho_q(w; x) \propto \frac{1}{w}\exp\left[-\frac{1}{q}\left(\frac{|x|}{w}\right	)^q \right]  \quad \text{ where }\; q > 0\;,
	\end{equation}
	which are known as the \emph{exponential power distributions} and include the Gaussian distribution ($q = 2$), the exponential distribution ($q = 1$) and the uniform distribution ($q \to \infty$).
	The integral over the width then becomes
	\begin{equation}
		\begin{split}
			P(\vec{x}) \propto &\int_{0}^{\infty} dw \ \frac{1}{w^N} \exp\left[-\sum_i \frac{1}{q}\left(\frac{|x_i|}{w}\right	)^q \right] \\
			= &\int_{0}^{\infty} dw \ \frac{1}{w^N} \exp\left[- \frac{1}{q}\frac{\|\vec{x}\|_q^q}{w^q} \right]\;,
		\end{split}
	\end{equation}
	where the $L_q$-norm~\footnote{Strictly speaking, this expression does not define a norm when $q<1$ because it violates the triangle inequality. Nevertheless, our  results still hold even in that case. } is defined by
	\begin{equation}
		\|\vec{x}\|_q \coloneqq \left(\sum_i |x_i|^q\right)^{1/q}\;.
	\end{equation}
	We use this norm to make the following change of variable 
	\begin{equation}
		z \coloneqq \frac{\|\vec{x}\|_q}{w} \Rightarrow \frac{dw}{dz} = -\frac{\|\vec{x}\|_q}{z^2}\;,
	\end{equation}
	so that the integral over the new variable $z$ becomes independent of the grid points, leaving us with
	\begin{equation}\label{eq:estochs_prior}
		P(\vec{x}) \propto \frac{1}{\|\vec{x}\|_q^{N-1}}\;.
	\end{equation}
	From this, we see that weighting the grid points vector by a power of its $L_q$-norm is equivalent to using a $q$-th power exponential function as a grid density and integrating flat over the width parameter.
	More specifically, using the $L_2$-norm is equivalent to using a Gaussian grid density and integrating over its standard deviation.
	Similarly, using the $L_1$-norm is equivalent to using an exponential density and integrating over the scale parameter while using the $L_\infty$-norm corresponds to using a uniform density and integrating over the cutoff. 
	Note that using a reciprocal prior in the width integral, as appropriate for scale parameters 
	\cite[p.~109]{Sivia06}, instead of a flat one leads to a very similar result where the power of the norm is $N{-}2$ instead of $N{-}1$. The difference between the two choices is insignificant in practice.
	
	Substituting Eq.~\eqref{eq:estochs_prior} back into the original expression Eq.~\eqref{eq:width_avg} and absorbing the integral over $z$ in the overall normalization constant we get the formal expression for this extended version of the average spectrum method
	\begin{equation}\label{eq:estochs}
		f_\text{ASM}(\rho_q, N; x) \!\propto\! 
		\!\int\!  d\vec{x}\, \frac{1}{\|\vec{x}\|_q^{N-1}} \!\int\limits_{0}^{\infty}\!  d\vec{\bar{f}} \, \mathrm{e}^{-\frac12\chi^2[\vec{\bar{f}}, \vec{x}]}\, f(\vec{\bar{f}},\vec{x}; x).
	\end{equation}
	We can easily adapt the sampling algorithm for released-grid ASM discussed in Sec.~\ref{sec:sampling} to evaluate this expression:
	We simply replace the prior density function in the acceptance ratio by the power ratio of the norms of grid samples. We only need to choose a reasonable value for the width parameter $w$ of the proposal distribution which can be easily estimated, e.g., from the non-negative least squares (NNLS) solution~\cite{Lawson95}.
	This value need not be very close to the width of the exact model. We have run calculations where it was an order of magnitude off. Obviously, its choice only affects the acceptance ratio of grid point sampling.
	
	\begin{figure}
		\center
		\includegraphics[width=0.95\columnwidth]{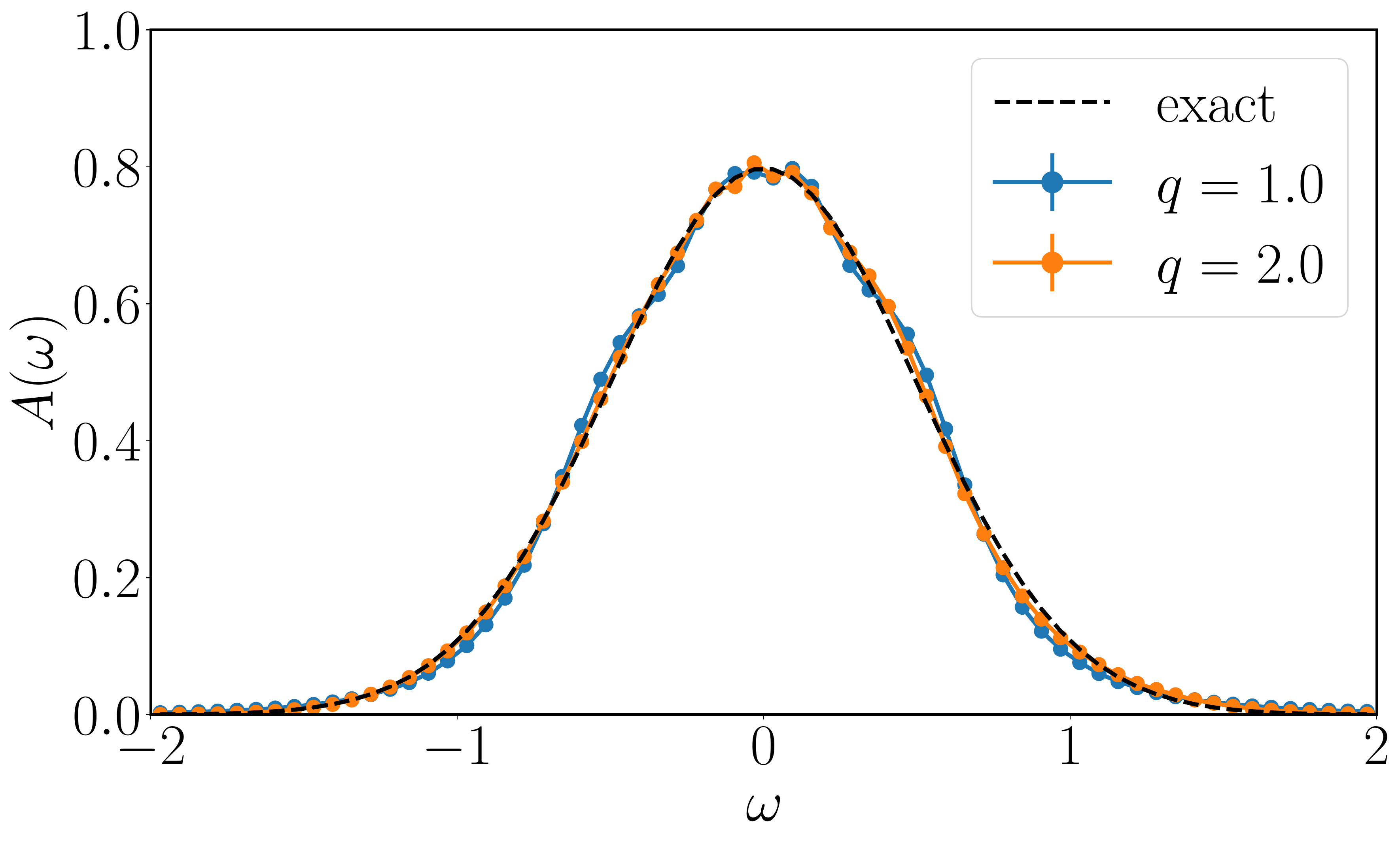}
		\caption{\label{fig:estochs}
			Spectral function $A(\omega)$ obtained using width-sampling ASM
			for the same problem as in Figs.~\ref{fig:width} and \ref{fig:maxent}.
			An exponential ($q=1$)  and a  Gaussian grid density ($q=2$) with $512$ grid points is used.
			The samples are binned on a uniform grid.
		}
	\end{figure}
	
	\begin{figure}
		\center
		\includegraphics[width=0.95\columnwidth]{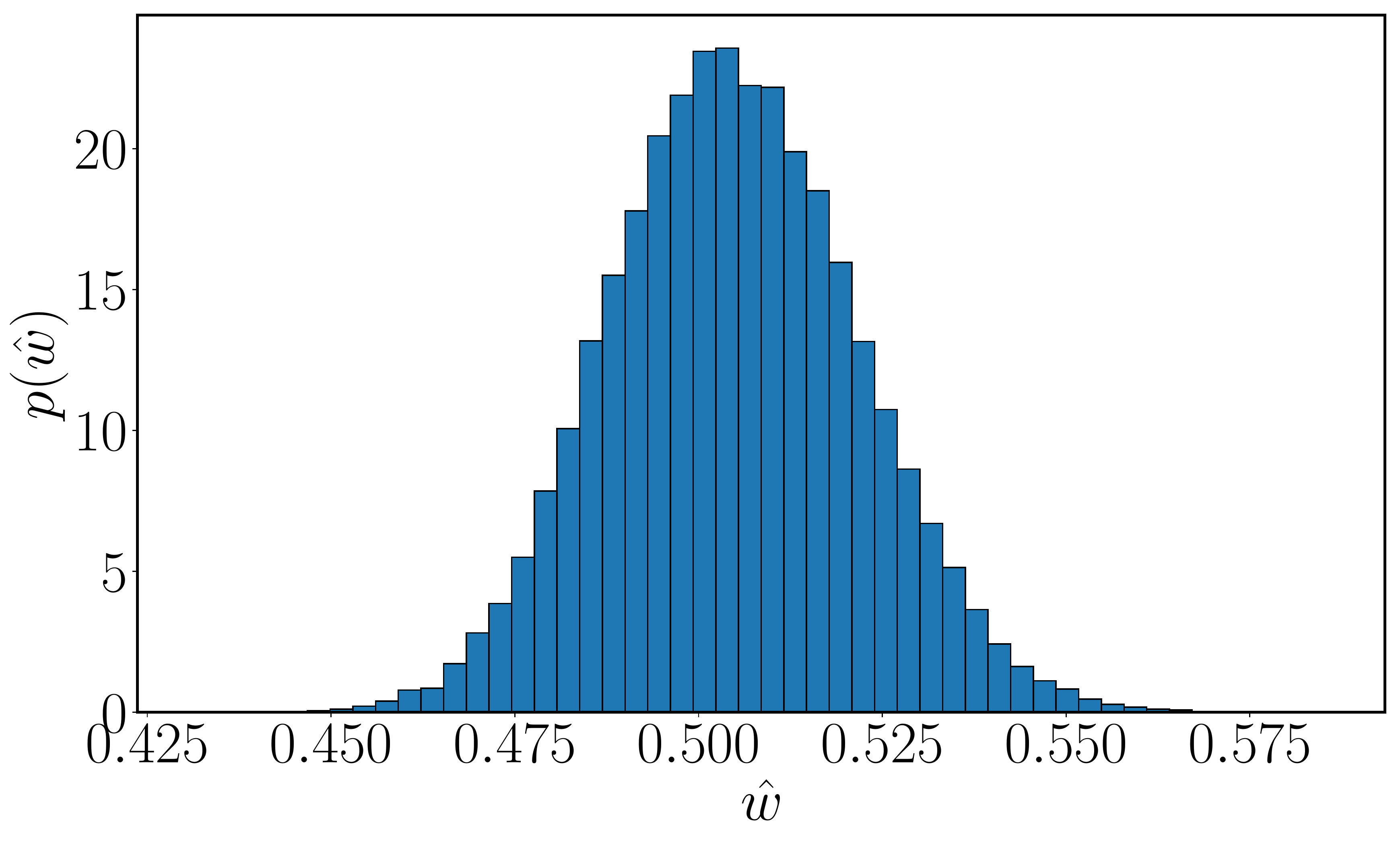}
		\caption{\label{fig:L2}
			Histogram of the scaled $L_2$-norm of grids sampled for the case $q=2$ of Fig.~\ref{fig:estochs}. 
			The scaled $L_2$-norm of a grid $\vec{x}$ is calculated as the standard deviation of its points $\hat{w} \coloneqq \sqrt[2]{{\|\vec{x}\|_2^2}/{N}} = \sqrt{{\sum_i x_i^2}/{N}}$.
			The widths are centered around the width of the exact model ($0.5)$.
		}
	\end{figure}
	
	\begin{figure}
		\center
		\includegraphics[width=0.95\columnwidth]{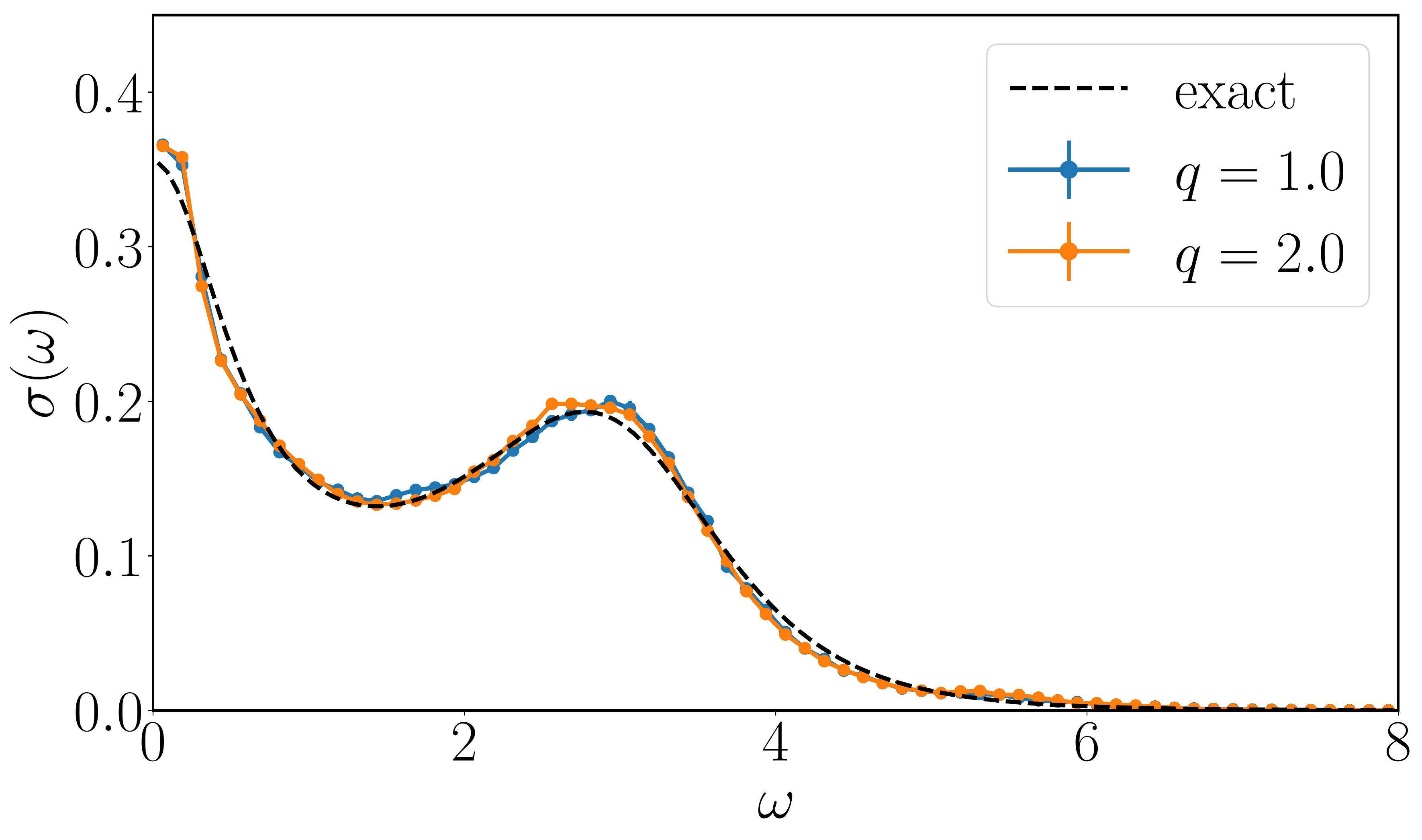}
		\caption{\label{fig:estochs_cutoff}
			Optical conductivity $\sigma(\omega)$ obtained using width-sampling ASM for the same problem as in Figs.~\ref{fig:cutoff} and \ref{fig:density}.
			An exponential ($q=1$)  and a  Gaussian grid density ($q=2$) with $512$ grid points is used.
			The samples are binned on a uniform grid.		
		}
	\end{figure}
	
	In Fig.~\ref{fig:estochs}, we show the results of width-sampling ASM for the same test case as in Fig.~\ref{fig:width}. 
	We did the calculations using both exponential ($L_1$-norm) and Gaussian grid densities ($L_2$-norm).
	The results of both calculations are in excellent agreement with the exact spectrum without the need for fine-tuning the exact value of the width parameter. Width-sampling ASM produces even better agreement than MaxEnt (Fig.~\ref{fig:maxent}).
	
	In Fig.~\ref{fig:L2},  we show the histogram of the scaled $L_2$-norm of grid samples from Gaussian densities ($q=2$).
	We calculated the scaled norm of a grid sample as $\hat{w} \coloneqq \sqrt{{\sum_i x_i^2}/{N}}$. 
	Notice how the method automatically finds the optimal value of 0.5 and averages around it to the extent allowed by the noise in the data.
	
	For completeness, we also report the effect of width averaging on the optical conductivity test case. The earlier results with released grid points were already very good and showed no noticable dependence on the width, so it does not come as a surprise that the results with width averaging are as good, as can be seen from Fig.~\ref{fig:estochs_cutoff}.
	
	Note that, unlike the fixed- and released-grid average spectrum method, width-sampling ASM does not have a default model: in the absence of data, except for a sum rule, the method does not give a result. This is by design: For convergence, width-averaging requires the data to provide information about the width. Having the least informative prior, width-sampling ASM thus is the least biased of the methods discussed here as may be seen by comparing Figs.~\ref{fig:width} and \ref{fig:maxent} with Fig.~\ref{fig:estochs}.
	
	\section{Grid Size Dependence}
\begin{figure}
		\center
		\includegraphics[width=0.95\columnwidth]{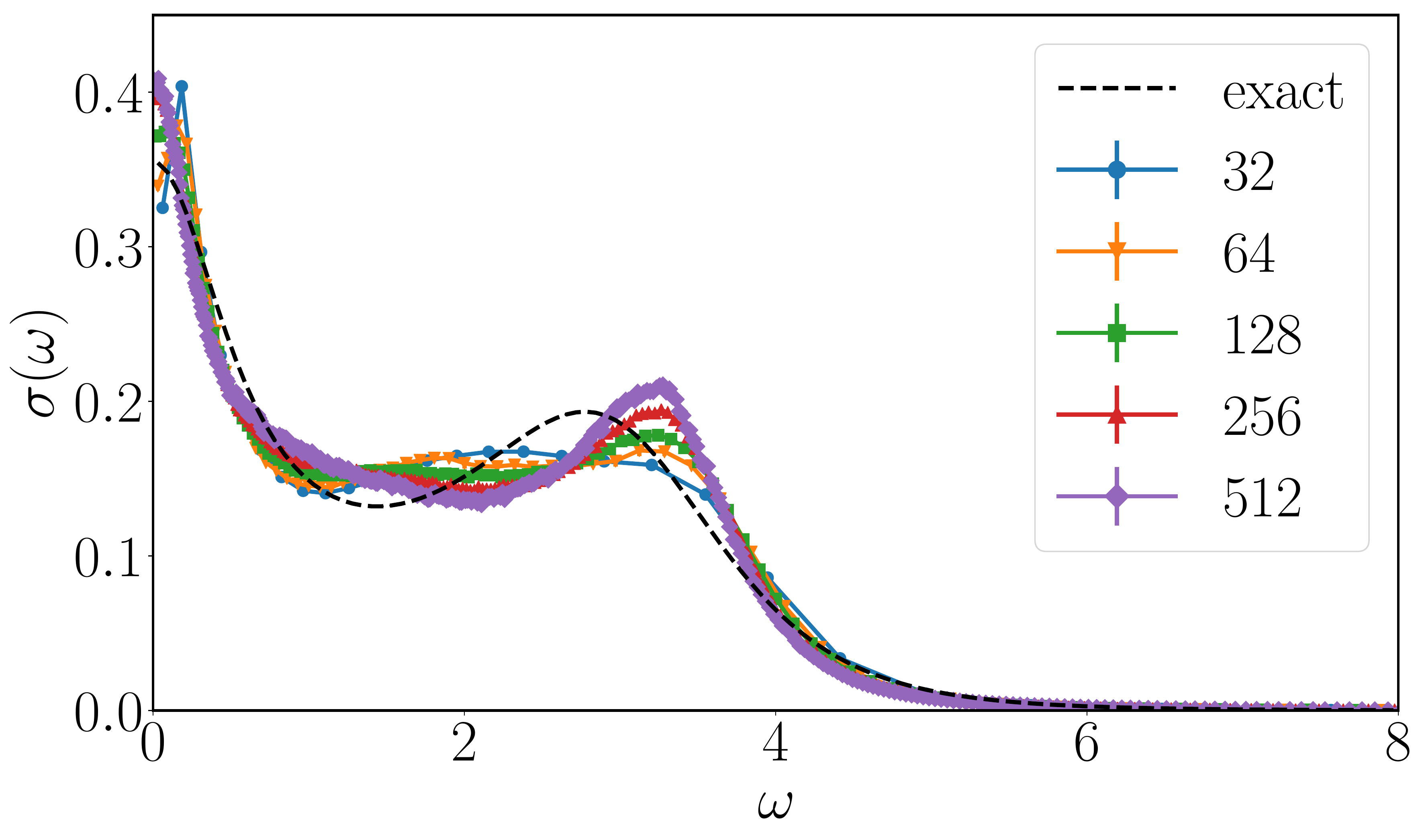}
		\includegraphics[width=0.95\columnwidth]{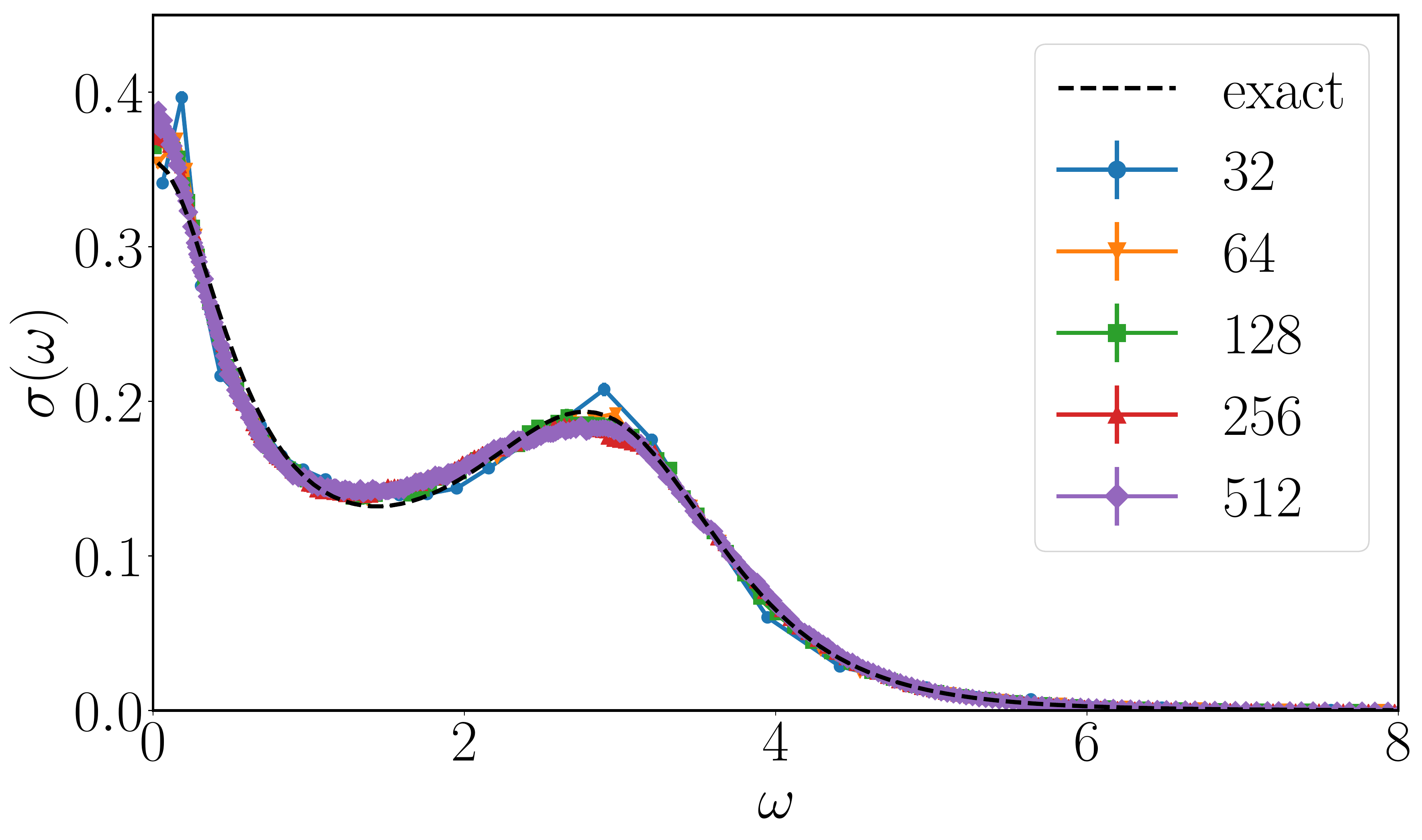}
		\caption{\label{fig:size}
			Dependence of the optical conductivity $\sigma(\omega)$ obtained using fixed-grid ASM (top) and released-grid ASM (bottom) on the grid size $N$ (label).
			A Lorentzian grid density with parameter $\gamma=2.5$ is used.
			For ease of comparison, the samples of the released calculations are binned and averaged on the grids of the corresponding fixed calculations.		
		}
	\end{figure}

	In ASM1, we discussed the dependence of the fixed-grid average spectrum method on  the number of grid points and found that it plays the role of a regularization parameter:
	as the number of grid points increases,  the results change and get more biased towards the grid density.
	The same behavior still applies to the released-grid ASM but is much weaker. This is in line with the overall reduction in the dependence on the grid density we saw earlier.
	For example, we compare in Fig.~\ref{fig:size} the $N$-dependence of the optical conductivity test case using fixed and released grid points. 
	While the fixed-grid calculations show a significant variation, released-grid ASM results hardly change with the number of grid points except for small differences near $\omega=0$. 
	We know that using a much larger number of grid points, even the released-grid method will eventually show a more pronounced dependence on the grid size. 
	We did not, however, observe this dependence in this test case for any reasonable value of $N$.
	To see it in released-gird ASM or its width-sampling extension, we need to look at yet another case where the exact spectrum is significantly different from the singly-peaked default models we typically use.
	
	To this end, let us take a spectral function composed of four Gaussian peaks symmetric around zero. 
	Two of the peaks are narrow with width $0.1$ and weight $0.15$ and located at frequencies $\pm0.5$. 
	The other two are wider with width $0.5$ and weight $0.35$ and located further out at $\pm2$.
	As with the previous spectral function, we generate the green function data using Eq.~\eqref{eq:tau_spectral} on 60 $\tau$-points for $\beta=50$.
	We add relative Gaussian noise with a standard deviation of $10^{-2}$. 
	
	\begin{figure}
		\center
		\includegraphics[width=0.95\columnwidth]{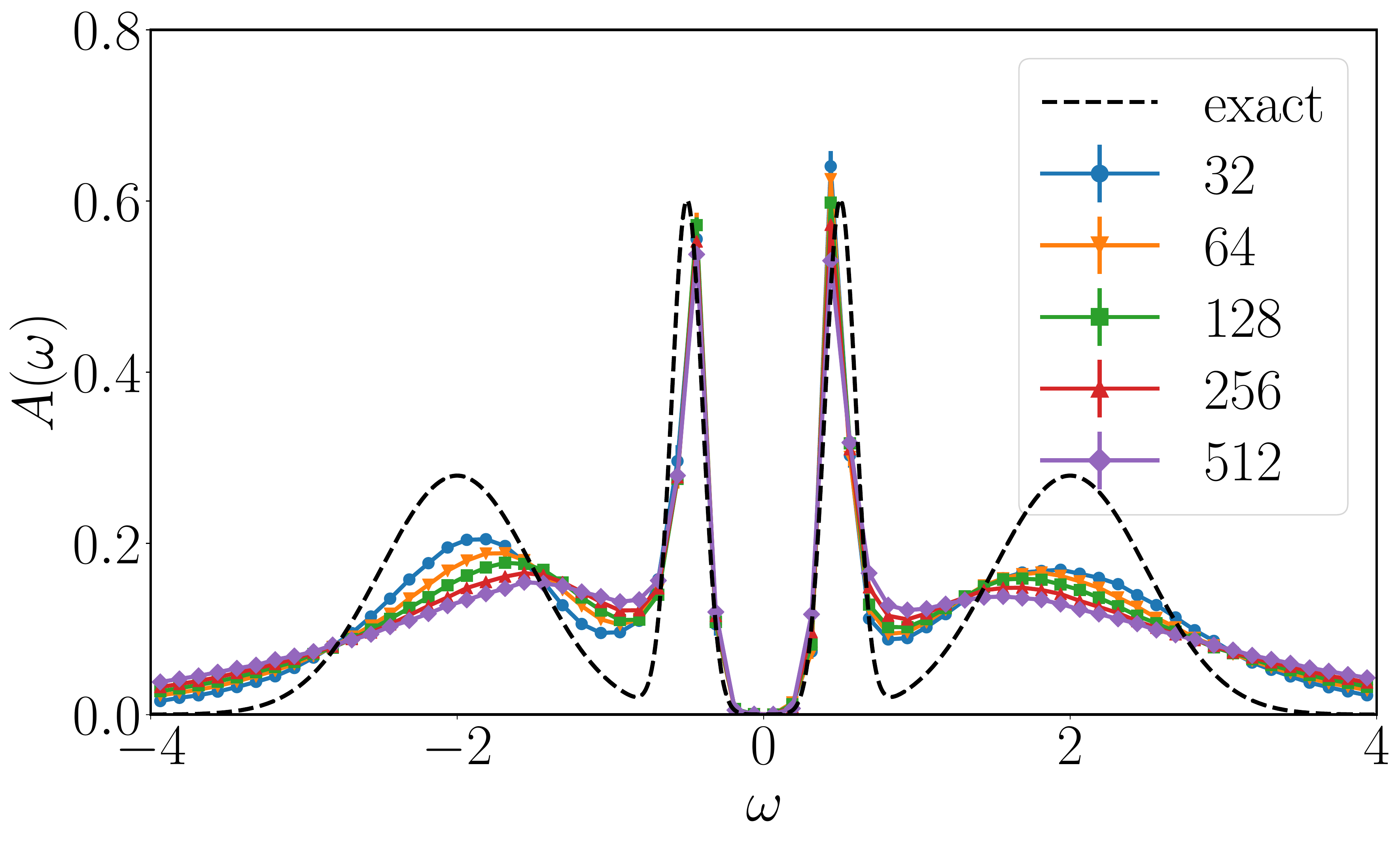}
		\includegraphics[width=0.95\columnwidth]{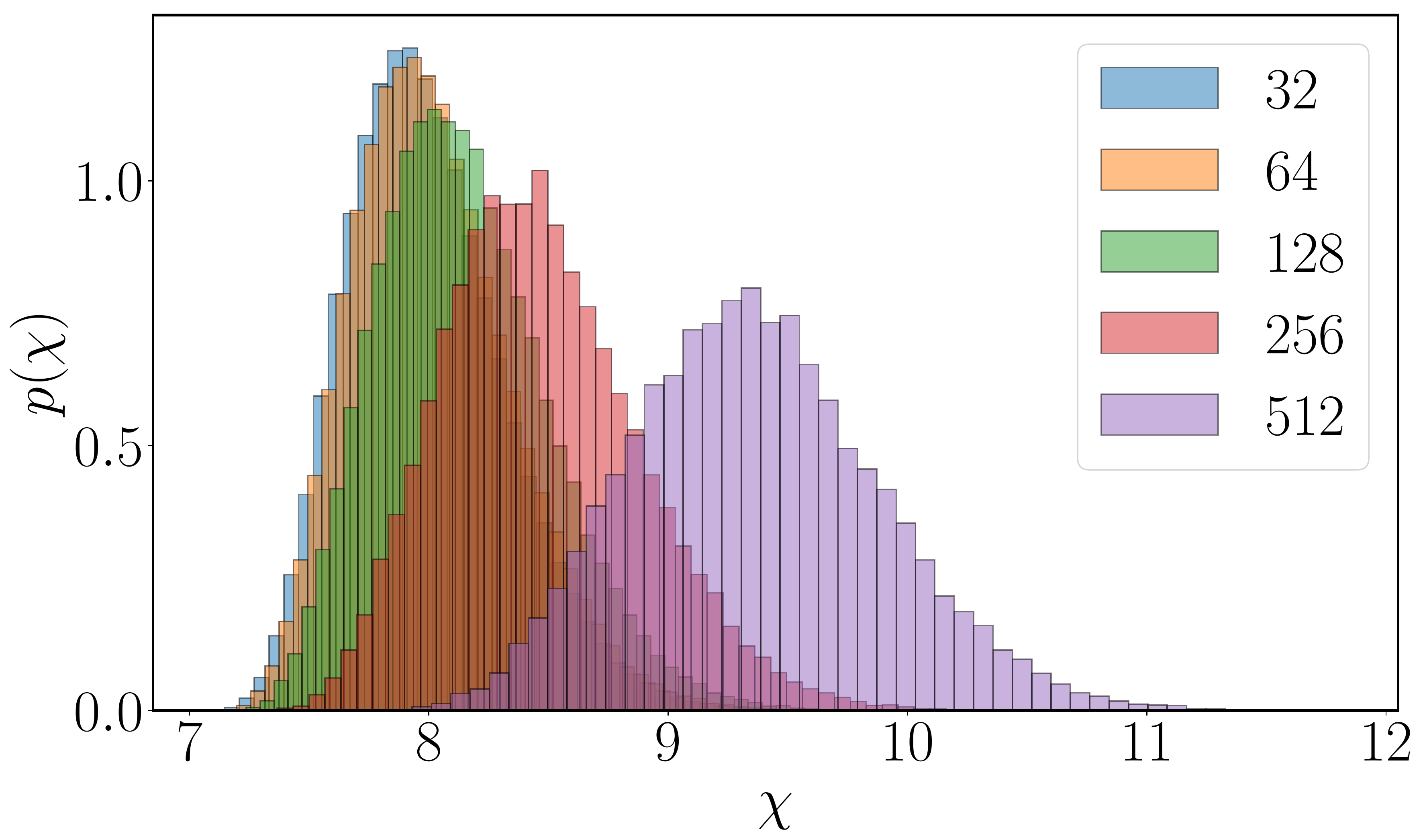}
		\caption{\label{fig:four_peaks}
			Grid size dependence of the spectral function $A(\omega)$ of 4 peaks using width-sampling ASM.
			A Gaussian grid density ($q=2$) is used.
			The samples are binned on a uniform grid.
			The bottom panel shows histograms of the data fits of sampled spectra.
		}
	\end{figure}
	
	\begin{figure}
		\center
		\includegraphics[width=0.95\columnwidth]{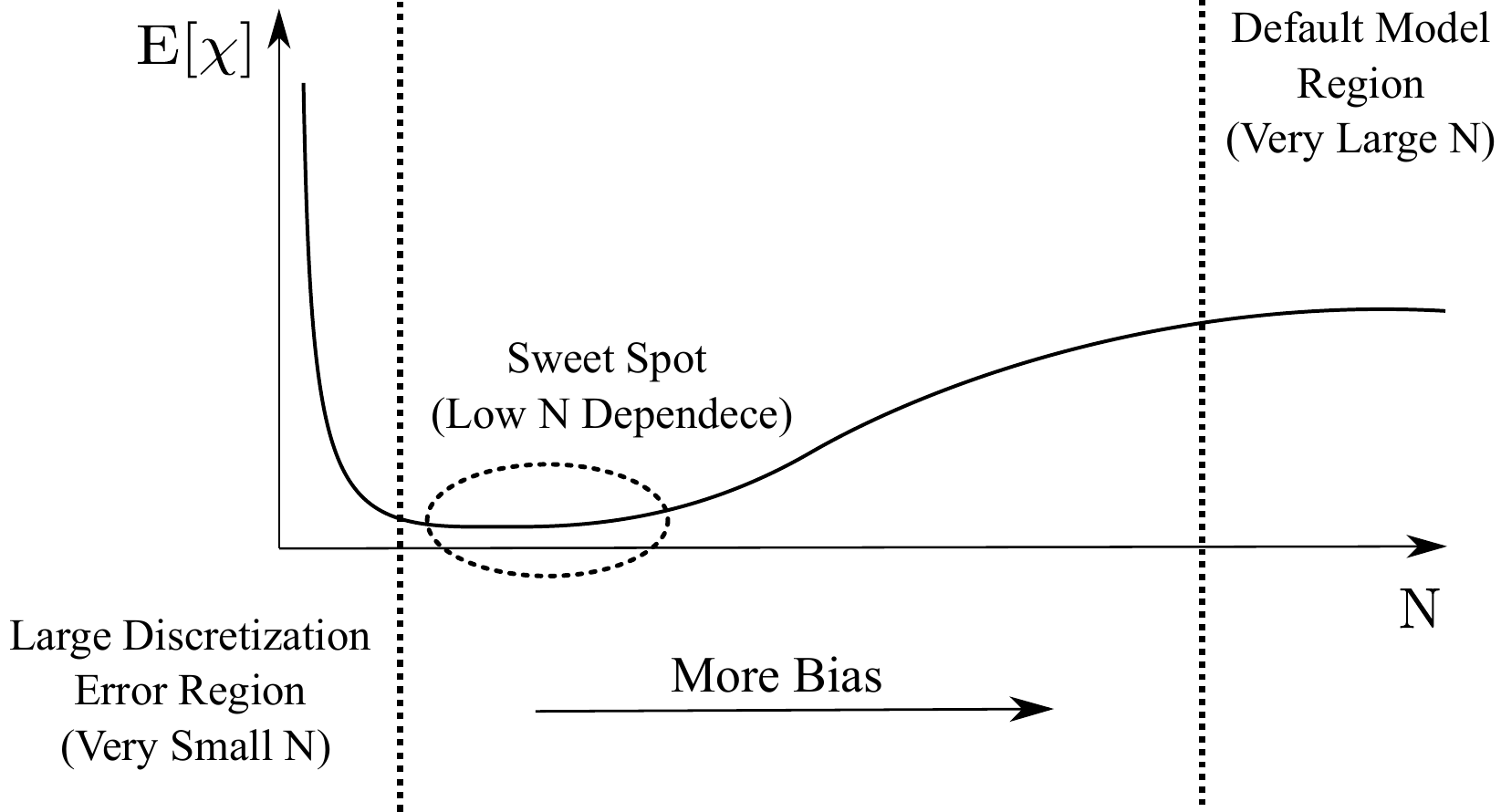}
		\caption{\label{fig:fit_size}
			Schematic diagram of the average fit behavior in ASM  as the number of grid points $N$ changes.
		}
	\end{figure}
	
	In the top panel of Fig.~\ref{fig:four_peaks}, we plot the results of width-sampling ASM using a Gaussian grid density ($q=2$) and different grid sizes.
	As the grid size increases, the results get smoother and the peaks get wider demonstrating the regularizing effect of the grid size even in the released-grid case.
	The behavior is similar using a fixed width.
	A better understanding is gained by checking the fit histograms of the sampled spectra in the bottom panel.
	As the grid size increases beyond $N=128$, the fit histograms, introduced in ASM1, shift to the right and get wider showing a systematic bias towards spectra of worse fits.
	On the other end, we observed that when the grid size is lower than $N=16$, the data fit gets extremely bad due to the large discretization error. We did not include results using such low grid sizes to avoid cluttering the plots.
	
	The qualitative behavior of the fit in the average spectrum methods is depicted in Fig.~\ref{fig:fit_size}.
	For very low grid sizes, the discretization error dominates, leading to a very bad fit. 
	Once the grid size is large enough such that the discretization error becomes negligible relative to the noise on the data, increasing the grid size leads to more bias and worse fit.
	This dependence on the grid size starts out slowly and then accelerates till the average spectrum approaches the default model (grid density) for very large $N$.
	Therefore, a weak dependence on the grid size indicates a better grid density, and when the density matches the exact model, the sweet spot extends to infinity.
	This behavior applies equally to both the fixed and the released-grid ASM.
	However, there are two main differences.
	The \emph{discretization} error for released-gird ASM is normally less than that of the fixed grid.
	Also, the region of low $N$-dependence extends further because the bias towards the grid density is much reduced compared to fixed-grid ASM.
	In general, the results of released-grid ASM show weak dependence for the typical grid sizes we use.
	A good recipe for choosing $N$ is to use the largest value for which the fit does not get substantially worse.
	For the test case of Fig.~\ref{fig:four_peaks} this would be $N=128$.
	
	It may be worth mentioning that we tried sampling the grid size using a flat prior.
	We found  that the method chooses a high number of points when the default model is highly compatible with the data as in Fig.~\ref{fig:size}.
	But when the default model does not fit the data very well, the method chooses a very low number of points.
	This happened in cases like the one shown in Fig.~\ref{fig:four_peaks}.
	In the end, we decided to keep the grid size as an independent parameter of the method that we use for checking the reliability of the results:
	A strong dependence on the grid size indicates a strong bias towards the default model and away from data.
	
	\section{Summary and Discussion}
	The results of the average spectrum method can strongly depend on the discretization grid.
	One approach for handling this dependence is choosing a grid density that fits the data decently.
	We showed that sampling the grid points helps to reduce the bias dramatically and thus obviates, in many cases, the need to search for the best grid.
	But in some cases, the width of the grid density still influences the results noticeably.
	We, therefore, went one step further and averaged over the results for grids of different widths.
	Remarkably, for a large family of grid densities, we could perform this additional sampling analytically, incurring only a negligible computational overhead.
	
	The approach used here for handling the grid dependence led us to the following hierarchy of average spectrum methods where each method extends the previous one by averaging over the relevant parameters
	\begin{equation}
		\underbrace{\int\! dw  \underbrace{\int\! d\vec{x} \prod_{i=1}^{N} \rho_q(w; x_i) \underbrace{\int\limits_{0}^{\infty}\!  d\vec{\bar{f}} \; \mathrm{e}^{-\frac12\chi^2[\vec{\bar{f}}, \vec{x}]}\, f(\vec{\bar{f}},\vec{x}; x)}_{f_\text{ASM}(\vec{x}; x)}}_{f_\text{ASM}(\rho_q, w, N; x)}}_{f_\text{ASM}(\rho_q, N; x)} .
	\end{equation}
	It is obvious that the approach can be taken further
	by varying over other parameters of the grid density.
	Had we, for example, observed a strong dependence on the functional form of the grid density, e.g. the parameter $q$ in Eq.~\eqref{eq:estochs}, we would have average the results of different such values using suitable weights.
	
	From this perspective,  the functional-integral based average spectrum method can be seen as a general framework for obtaining data-compatible spectra in the context of analytic continuation and similar spectral reconstruction problems.
	Given a certain parametrization of the spectrum, the most straightforward solution is estimating these parameters by fitting the data.
	In many situations, this may be enough to single out a small region of the parameter space with tolerable variations in the spectrum.
	When observing a noticeable sensitivity of the results to a parameter, one should consider averaging over the results of different values of this parameter to smooth out details not supported by the data.
	In light of this, the fixed-gird method of ASM1 can be seen as an extension of the non-negative least squares method (NNLS), where instead of finding the spectral integrals that fits the data best, it averages over, giving equal weights to all spectral integrals fitting the data equally.
	
	In this paper, we could discuss only a few test cases, each of which was introduced for illustrating specific aspects of the average spectrum methods.
	Further test cases and applications of theses extensions to real-world problems are discussed in Ref.~\cite{Ghanem17}.
	To encourage further testing of the ASM methods and their use for practical problems, efficient web-based implementations of the different flavors of ASM are accessible at~\cite{Spektra}.

\providecommand{\noopsort}[1]{}\providecommand{\singleletter}[1]{#1}%


\begin{thebibliography}{28}%
	\makeatletter
	\providecommand \@ifxundefined [1]{%
		\@ifx{#1\undefined}
	}%
	\providecommand \@ifnum [1]{%
		\ifnum #1\expandafter \@firstoftwo
		\else \expandafter \@secondoftwo
		\fi
	}%
	\providecommand \@ifx [1]{%
		\ifx #1\expandafter \@firstoftwo
		\else \expandafter \@secondoftwo
		\fi
	}%
	\providecommand \natexlab [1]{#1}%
	\providecommand \enquote  [1]{``#1''}%
	\providecommand \bibnamefont  [1]{#1}%
	\providecommand \bibfnamefont [1]{#1}%
	\providecommand \citenamefont [1]{#1}%
	\providecommand \href@noop [0]{\@secondoftwo}%
	\providecommand \href [0]{\begingroup \@sanitize@url \@href}%
	\providecommand \@href[1]{\@@startlink{#1}\@@href}%
	\providecommand \@@href[1]{\endgroup#1\@@endlink}%
	\providecommand \@sanitize@url [0]{\catcode `\\12\catcode `\$12\catcode
		`\&12\catcode `\#12\catcode `\^12\catcode `\_12\catcode `\%12\relax}%
	\providecommand \@@startlink[1]{}%
	\providecommand \@@endlink[0]{}%
	\providecommand \url  [0]{\begingroup\@sanitize@url \@url }%
	\providecommand \@url [1]{\endgroup\@href {#1}{\urlprefix }}%
	\providecommand \urlprefix  [0]{URL }%
	\providecommand \Eprint [0]{\href }%
	\providecommand \doibase [0]{http://dx.doi.org/}%
	\providecommand \selectlanguage [0]{\@gobble}%
	\providecommand \bibinfo  [0]{\@secondoftwo}%
	\providecommand \bibfield  [0]{\@secondoftwo}%
	\providecommand \translation [1]{[#1]}%
	\providecommand \BibitemOpen [0]{}%
	\providecommand \bibitemStop [0]{}%
	\providecommand \bibitemNoStop [0]{.\EOS\space}%
	\providecommand \EOS [0]{\spacefactor3000\relax}%
	\providecommand \BibitemShut  [1]{\csname bibitem#1\endcsname}%
	\let\auto@bib@innerbib\@empty
	\bibitem [{\citenamefont {Silver}\ \emph {et~al.}(1990)\citenamefont {Silver},
		\citenamefont {Sivia},\ and\ \citenamefont {Gubernatis}}]{Silver90}%
	\BibitemOpen
	\bibfield  {author} {\bibinfo {author} {\bibfnamefont {R.~N.}\ \bibnamefont
			{Silver}}, \bibinfo {author} {\bibfnamefont {D.~S.}\ \bibnamefont {Sivia}}, \
		and\ \bibinfo {author} {\bibfnamefont {J.~E.}\ \bibnamefont {Gubernatis}},\
	}\href {\doibase 10.1103/PhysRevB.41.2380} {\bibfield  {journal} {\bibinfo
			{journal} {Phys. Rev. B}\ }\textbf {\bibinfo {volume} {41}},\ \bibinfo
		{pages} {2380} (\bibinfo {year} {1990})}\BibitemShut {NoStop}%
	\bibitem [{\citenamefont {Jarrell}\ and\ \citenamefont
		{Gubernatis}(1996)}]{Jarrell96}%
	\BibitemOpen
	\bibfield  {author} {\bibinfo {author} {\bibfnamefont {M.}~\bibnamefont
			{Jarrell}}\ and\ \bibinfo {author} {\bibfnamefont {J.~E.}\ \bibnamefont
			{Gubernatis}},\ }\href {\doibase 10.1016/0370-1573(95)00074-7} {\bibfield
		{journal} {\bibinfo  {journal} {Phys. Rep.}\ }\textbf {\bibinfo {volume}
			{269}},\ \bibinfo {pages} {133} (\bibinfo {year} {1996})}\BibitemShut
	{NoStop}%
	\bibitem [{\citenamefont {Gunnarsson}\ \emph
		{et~al.}(2010{\natexlab{a}})\citenamefont {Gunnarsson}, \citenamefont
		{Haverkort},\ and\ \citenamefont {Sangiovanni}}]{Gunnarsson10}%
	\BibitemOpen
	\bibfield  {author} {\bibinfo {author} {\bibfnamefont {O.}~\bibnamefont
			{Gunnarsson}}, \bibinfo {author} {\bibfnamefont {M.~W.}\ \bibnamefont
			{Haverkort}}, \ and\ \bibinfo {author} {\bibfnamefont {G.}~\bibnamefont
			{Sangiovanni}},\ }\href {\doibase 10.1103/PhysRevB.81.155107} {\bibfield
		{journal} {\bibinfo  {journal} {Phys. Rev. B}\ }\textbf {\bibinfo {volume}
			{81}},\ \bibinfo {pages} {155107} (\bibinfo {year}
		{2010}{\natexlab{a}})}\BibitemShut {NoStop}%
	\bibitem [{\citenamefont {Jarrell}(2012)}]{Jarrell12}%
	\BibitemOpen
	\bibfield  {author} {\bibinfo {author} {\bibfnamefont {M.}~\bibnamefont
			{Jarrell}},\ }in\ \href {http://www.cond-mat.de/events/correl12/manuscripts}
	{\emph {\bibinfo {booktitle} {Correlated Electrons: From Models to
				Materials}}},\ \bibinfo {editor} {edited by\ \bibinfo {editor} {\bibfnamefont
			{E.}~\bibnamefont {Pavarini}}, \bibinfo {editor} {\bibfnamefont
			{E.}~\bibnamefont {Koch}}, \bibinfo {editor} {\bibfnamefont {F.}~\bibnamefont
			{Anders}}, \ and\ \bibinfo {editor} {\bibfnamefont {M.}~\bibnamefont
			{Jarrell}}}\ (\bibinfo  {publisher} {Forschungszentrum J\"ulich},\ \bibinfo
	{address} {J\"ulich},\ \bibinfo {year} {2012})\BibitemShut {NoStop}%
	\bibitem [{\citenamefont {Reymbaut}\ \emph {et~al.}(2015)\citenamefont
		{Reymbaut}, \citenamefont {Bergeron},\ and\ \citenamefont
		{Tremblay}}]{Tremblay15}%
	\BibitemOpen
	\bibfield  {author} {\bibinfo {author} {\bibfnamefont {A.}~\bibnamefont
			{Reymbaut}}, \bibinfo {author} {\bibfnamefont {D.}~\bibnamefont {Bergeron}},
		\ and\ \bibinfo {author} {\bibfnamefont {A.-M.~S.}\ \bibnamefont
			{Tremblay}},\ }\href {\doibase 10.1103/PhysRevB.92.060509} {\bibfield
		{journal} {\bibinfo  {journal} {Phys. Rev. B}\ }\textbf {\bibinfo {volume}
			{92}},\ \bibinfo {pages} {060509(R)} (\bibinfo {year} {2015})}\BibitemShut
	{NoStop}%
	\bibitem [{\citenamefont {White}(1991)}]{White91}%
	\BibitemOpen
	\bibfield  {author} {\bibinfo {author} {\bibfnamefont {S.~R.}\ \bibnamefont
			{White}},\ }in\ \href {\doibase 10.1007/978-3-642-76382-3_13} {\emph
		{\bibinfo {booktitle} {Computer Simulation Studies in Condensed Matter
				Physics {III}}}},\ \bibinfo {editor} {edited by\ \bibinfo {editor}
		{\bibfnamefont {D.~P.}\ \bibnamefont {Landau}}, \bibinfo {editor}
		{\bibfnamefont {K.~K.}\ \bibnamefont {Mon}}, \ and\ \bibinfo {editor}
		{\bibfnamefont {B.-B.}\ \bibnamefont {Sch\"uttler}}}\ (\bibinfo  {publisher}
	{Springer},\ \bibinfo {address} {Heidelberg},\ \bibinfo {year} {1991})\ pp.\
	\bibinfo {pages} {145--153}\BibitemShut {NoStop}%
	\bibitem [{\citenamefont {Sandvik}(1998)}]{Sandvik98}%
	\BibitemOpen
	\bibfield  {author} {\bibinfo {author} {\bibfnamefont {A.~W.}\ \bibnamefont
			{Sandvik}},\ }\href {\doibase 10.1103/PhysRevB.57.10287} {\bibfield
		{journal} {\bibinfo  {journal} {Phys. Rev. B}\ }\textbf {\bibinfo {volume}
			{57}},\ \bibinfo {pages} {10287} (\bibinfo {year} {1998})}\BibitemShut
	{NoStop}%
	\bibitem [{\citenamefont {Beach}(2004)}]{Beach04}%
	\BibitemOpen
	\bibfield  {author} {\bibinfo {author} {\bibfnamefont {K.~S.~D.}\
			\bibnamefont {Beach}},\ }\href {https://arxiv.org/abs/cond-mat/0403055}
	{\enquote {\bibinfo {title} {Identifying the maximum entropy method as a
				special limit of stochastic analytic continuation},}\ } (\bibinfo {year}
	{2004}),\ \Eprint {http://arxiv.org/abs/arXiv:cond-mat/0403055}
	{arXiv:cond-mat/0403055} \BibitemShut {NoStop}%
	\bibitem [{\citenamefont {Vafayi}\ and\ \citenamefont
		{Gunnarsson}(2007)}]{Gunnarsson07}%
	\BibitemOpen
	\bibfield  {author} {\bibinfo {author} {\bibfnamefont {K.}~\bibnamefont
			{Vafayi}}\ and\ \bibinfo {author} {\bibfnamefont {O.}~\bibnamefont
			{Gunnarsson}},\ }\href {\doibase 10.1103/PhysRevB.76.035115} {\bibfield
		{journal} {\bibinfo  {journal} {Phys. Rev. B}\ }\textbf {\bibinfo {volume}
			{76}},\ \bibinfo {pages} {035115} (\bibinfo {year} {2007})}\BibitemShut
	{NoStop}%
	\bibitem [{\citenamefont {Sylju\r{a}sen}(2008)}]{Syljuasen08}%
	\BibitemOpen
	\bibfield  {author} {\bibinfo {author} {\bibfnamefont {O.~F.}\ \bibnamefont
			{Sylju\r{a}sen}},\ }\href {\doibase 10.1103/physRevB.78.174429} {\bibfield
		{journal} {\bibinfo  {journal} {Phys. Rev. B}\ }\textbf {\bibinfo {volume}
			{78}},\ \bibinfo {pages} {174429} (\bibinfo {year} {2008})}\BibitemShut
	{NoStop}%
	\bibitem [{\citenamefont {Fuchs}\ \emph {et~al.}(2010)\citenamefont {Fuchs},
		\citenamefont {Pruschke},\ and\ \citenamefont {Jarrell}}]{Fuchs10}%
	\BibitemOpen
	\bibfield  {author} {\bibinfo {author} {\bibfnamefont {S.}~\bibnamefont
			{Fuchs}}, \bibinfo {author} {\bibfnamefont {T.}~\bibnamefont {Pruschke}}, \
		and\ \bibinfo {author} {\bibfnamefont {M.}~\bibnamefont {Jarrell}},\ }\href
	{\doibase 10.1103/PhysRevE.81.056701} {\bibfield  {journal} {\bibinfo
			{journal} {Phys. Rev. E}\ }\textbf {\bibinfo {volume} {81}},\ \bibinfo
		{pages} {056701} (\bibinfo {year} {2010})}\BibitemShut {NoStop}%
	\bibitem [{\citenamefont {Sandvik}(2016)}]{Sandvik16}%
	\BibitemOpen
	\bibfield  {author} {\bibinfo {author} {\bibfnamefont {A.~W.}\ \bibnamefont
			{Sandvik}},\ }\href {\doibase 10.1103/PhysRevE.94.063308} {\bibfield
		{journal} {\bibinfo  {journal} {Phys. Rev. E}\ }\textbf {\bibinfo {volume}
			{94}},\ \bibinfo {pages} {063308} (\bibinfo {year} {2016})}\BibitemShut
	{NoStop}%
	\bibitem [{\citenamefont {Ghanem}\ and\ \citenamefont {Koch}(2020)}]{Ghanem20}%
	\BibitemOpen
	\bibfield  {author} {\bibinfo {author} {\bibfnamefont {K.}~\bibnamefont
			{Ghanem}}\ and\ \bibinfo {author} {\bibfnamefont {E.}~\bibnamefont {Koch}},\
	}\href {\doibase 10.1103/PhysRevB.101.085111} {\bibfield  {journal} {\bibinfo
			{journal} {Phys. Rev. B}\ }\textbf {\bibinfo {volume} {101}},\ \bibinfo
		{pages} {085111} (\bibinfo {year} {2020})}\BibitemShut {NoStop}%
	\bibitem [{\citenamefont {{Vidberg}}\ and\ \citenamefont
		{{Serene}}(1977)}]{Vidberg77}%
	\BibitemOpen
	\bibfield  {author} {\bibinfo {author} {\bibfnamefont {H.~J.}\ \bibnamefont
			{{Vidberg}}}\ and\ \bibinfo {author} {\bibfnamefont {J.~W.}\ \bibnamefont
			{{Serene}}},\ }\href {\doibase 10.1007/BF00655090} {\bibfield  {journal}
		{\bibinfo  {journal} {Journal of Low Temperature Physics}\ }\textbf {\bibinfo
			{volume} {29}},\ \bibinfo {pages} {179} (\bibinfo {year} {1977})}\BibitemShut
	{NoStop}%
	\bibitem [{\citenamefont {Beach}\ \emph {et~al.}(2000)\citenamefont {Beach},
		\citenamefont {Gooding},\ and\ \citenamefont {Marsiglio}}]{Beach00}%
	\BibitemOpen
	\bibfield  {author} {\bibinfo {author} {\bibfnamefont {K.~S.~D.}\
			\bibnamefont {Beach}}, \bibinfo {author} {\bibfnamefont {R.~J.}\ \bibnamefont
			{Gooding}}, \ and\ \bibinfo {author} {\bibfnamefont {F.}~\bibnamefont
			{Marsiglio}},\ }\href {\doibase 10.1103/PhysRevB.61.5147} {\bibfield
		{journal} {\bibinfo  {journal} {Phys. Rev. B}\ }\textbf {\bibinfo {volume}
			{61}},\ \bibinfo {pages} {5147} (\bibinfo {year} {2000})}\BibitemShut
	{NoStop}%
	\bibitem [{\citenamefont {\"Ostlin}\ \emph {et~al.}(2012)\citenamefont
		{\"Ostlin}, \citenamefont {Chioncel},\ and\ \citenamefont
		{Vitos}}]{Ostlin12}%
	\BibitemOpen
	\bibfield  {author} {\bibinfo {author} {\bibfnamefont {A.}~\bibnamefont
			{\"Ostlin}}, \bibinfo {author} {\bibfnamefont {L.}~\bibnamefont {Chioncel}},
		\ and\ \bibinfo {author} {\bibfnamefont {L.}~\bibnamefont {Vitos}},\ }\href
	{\doibase 10.1103/PhysRevB.86.235107} {\bibfield  {journal} {\bibinfo
			{journal} {Phys. Rev. B}\ }\textbf {\bibinfo {volume} {86}},\ \bibinfo
		{pages} {235107} (\bibinfo {year} {2012})}\BibitemShut {NoStop}%
	\bibitem [{\citenamefont {Sch\"ott}\ \emph {et~al.}(2016)\citenamefont
		{Sch\"ott}, \citenamefont {Locht}, \citenamefont {Lundin}, \citenamefont
		{Gr\aa{}n\"as}, \citenamefont {Eriksson},\ and\ \citenamefont
		{Di~Marco}}]{Schot16}%
	\BibitemOpen
	\bibfield  {author} {\bibinfo {author} {\bibfnamefont {J.}~\bibnamefont
			{Sch\"ott}}, \bibinfo {author} {\bibfnamefont {I.~L.~M.}\ \bibnamefont
			{Locht}}, \bibinfo {author} {\bibfnamefont {E.}~\bibnamefont {Lundin}},
		\bibinfo {author} {\bibfnamefont {O.}~\bibnamefont {Gr\aa{}n\"as}}, \bibinfo
		{author} {\bibfnamefont {O.}~\bibnamefont {Eriksson}}, \ and\ \bibinfo
		{author} {\bibfnamefont {I.}~\bibnamefont {Di~Marco}},\ }\href {\doibase
		10.1103/PhysRevB.93.075104} {\bibfield  {journal} {\bibinfo  {journal} {Phys.
				Rev. B}\ }\textbf {\bibinfo {volume} {93}},\ \bibinfo {pages} {075104}
		(\bibinfo {year} {2016})}\BibitemShut {NoStop}%
	\bibitem [{\citenamefont {Arsenault}\ \emph {et~al.}(2014)\citenamefont
		{Arsenault}, \citenamefont {Lopez-Bezanilla}, \citenamefont {von
			Lilienfeld},\ and\ \citenamefont {Millis}}]{Arsenault14}%
	\BibitemOpen
	\bibfield  {author} {\bibinfo {author} {\bibfnamefont {L.-F.}\ \bibnamefont
			{Arsenault}}, \bibinfo {author} {\bibfnamefont {A.}~\bibnamefont
			{Lopez-Bezanilla}}, \bibinfo {author} {\bibfnamefont {O.~A.}\ \bibnamefont
			{von Lilienfeld}}, \ and\ \bibinfo {author} {\bibfnamefont {A.~J.}\
			\bibnamefont {Millis}},\ }\href {\doibase 10.1103/PhysRevB.90.155136}
	{\bibfield  {journal} {\bibinfo  {journal} {Phys. Rev. B}\ }\textbf {\bibinfo
			{volume} {90}},\ \bibinfo {pages} {155136} (\bibinfo {year}
		{2014})}\BibitemShut {NoStop}%
	\bibitem [{\citenamefont {Arsenault}\ \emph {et~al.}(2017)\citenamefont
		{Arsenault}, \citenamefont {Neuberg}, \citenamefont {Hannah},\ and\
		\citenamefont {Andrew J.~Millis}}]{Arsenault16}%
	\BibitemOpen
	\bibfield  {author} {\bibinfo {author} {\bibfnamefont {L.-F.}\ \bibnamefont
			{Arsenault}}, \bibinfo {author} {\bibfnamefont {R.}~\bibnamefont {Neuberg}},
		\bibinfo {author} {\bibfnamefont {L.~A.}\ \bibnamefont {Hannah}}, \ and\
		\bibinfo {author} {\bibfnamefont {A.~J.}\ \bibnamefont {Andrew J.~Millis}},\
	}\href {\doibase https://doi.org/10.1088/1361-6420/aa8d93} {\bibfield
		{journal} {\bibinfo  {journal} {Inverse Problems}\ }\textbf {\bibinfo
			{volume} {33}},\ \bibinfo {pages} {115007} (\bibinfo {year}
		{2017})}\BibitemShut {NoStop}%
	\bibitem [{\citenamefont {Yoon}\ \emph {et~al.}(2018)\citenamefont {Yoon},
		\citenamefont {Sim},\ and\ \citenamefont {Han}}]{Yoon18}%
	\BibitemOpen
	\bibfield  {author} {\bibinfo {author} {\bibfnamefont {H.}~\bibnamefont
			{Yoon}}, \bibinfo {author} {\bibfnamefont {J.-H.}\ \bibnamefont {Sim}}, \
		and\ \bibinfo {author} {\bibfnamefont {M.~J.}\ \bibnamefont {Han}},\ }\href
	{\doibase 10.1103/PhysRevB.98.245101} {\bibfield  {journal} {\bibinfo
			{journal} {Phys. Rev. B}\ }\textbf {\bibinfo {volume} {98}},\ \bibinfo
		{pages} {245101} (\bibinfo {year} {2018})}\BibitemShut {NoStop}%
	\bibitem [{\citenamefont {Fournier}\ \emph {et~al.}(2020)\citenamefont
		{Fournier}, \citenamefont {Wang}, \citenamefont {Yazyev},\ and\ \citenamefont
		{Wu}}]{Fournier20}%
	\BibitemOpen
	\bibfield  {author} {\bibinfo {author} {\bibfnamefont {R.}~\bibnamefont
			{Fournier}}, \bibinfo {author} {\bibfnamefont {L.}~\bibnamefont {Wang}},
		\bibinfo {author} {\bibfnamefont {O.~V.}\ \bibnamefont {Yazyev}}, \ and\
		\bibinfo {author} {\bibfnamefont {Q.}~\bibnamefont {Wu}},\ }\href {\doibase
		10.1103/PhysRevLett.124.056401} {\bibfield  {journal} {\bibinfo  {journal}
			{Phys. Rev. Lett.}\ }\textbf {\bibinfo {volume} {124}},\ \bibinfo {pages}
		{056401} (\bibinfo {year} {2020})}\BibitemShut {NoStop}%
	\bibitem [{\citenamefont {Lawson}\ and\ \citenamefont
		{Hanson}(1995)}]{Lawson95}%
	\BibitemOpen
	\bibfield  {author} {\bibinfo {author} {\bibfnamefont {C.~L.}\ \bibnamefont
			{Lawson}}\ and\ \bibinfo {author} {\bibfnamefont {R.~J.}\ \bibnamefont
			{Hanson}},\ }\href {\doibase 10.1137/1.9781611971217} {\emph {\bibinfo
			{title} {Solving Least Squares Problems}}}\ (\bibinfo  {publisher} {SIAM},\
	\bibinfo {address} {Philadelphia},\ \bibinfo {year} {1995})\BibitemShut
	{NoStop}%
	\bibitem [{\citenamefont {Gunnarsson}\ \emph
		{et~al.}(2010{\natexlab{b}})\citenamefont {Gunnarsson}, \citenamefont
		{Haverkort},\ and\ \citenamefont {Sangiovanni}}]{Gunnarsson10b}%
	\BibitemOpen
	\bibfield  {author} {\bibinfo {author} {\bibfnamefont {O.}~\bibnamefont
			{Gunnarsson}}, \bibinfo {author} {\bibfnamefont {M.~W.}\ \bibnamefont
			{Haverkort}}, \ and\ \bibinfo {author} {\bibfnamefont {G.}~\bibnamefont
			{Sangiovanni}},\ }\href {\doibase 10.1103/PhysRevB.82.165125} {\bibfield
		{journal} {\bibinfo  {journal} {Phys. Rev. B}\ }\textbf {\bibinfo {volume}
			{82}},\ \bibinfo {pages} {165125} (\bibinfo {year}
		{2010}{\natexlab{b}})}\BibitemShut {NoStop}%
	\bibitem [{\citenamefont {Levy}\ \emph {et~al.}(2017)\citenamefont {Levy},
		\citenamefont {LeBlanc},\ and\ \citenamefont {Gull}}]{Levy17}%
	\BibitemOpen
	\bibfield  {author} {\bibinfo {author} {\bibfnamefont {R.}~\bibnamefont
			{Levy}}, \bibinfo {author} {\bibfnamefont {J.}~\bibnamefont {LeBlanc}}, \
		and\ \bibinfo {author} {\bibfnamefont {E.}~\bibnamefont {Gull}},\ }\href
	{\doibase https://doi.org/10.1016/j.cpc.2017.01.018} {\bibfield  {journal}
		{\bibinfo  {journal} {Computer Physics Communications}\ }\textbf {\bibinfo
			{volume} {215}},\ \bibinfo {pages} {149 } (\bibinfo {year}
		{2017})}\BibitemShut {NoStop}%
	\bibitem [{Note1()}]{Note1}%
	\BibitemOpen
	\bibinfo {note} {Strictly speaking, this expression does not define a norm
		when $q<1$ because it violates the triangle inequality. Nevertheless, our
		results still hold even in that case.}\BibitemShut {Stop}%
	\bibitem [{\citenamefont {Sivia}\ and\ \citenamefont
		{Skilling}(2006)}]{Sivia06}%
	\BibitemOpen
	\bibfield  {author} {\bibinfo {author} {\bibfnamefont {C.~S.}\ \bibnamefont
			{Sivia}}\ and\ \bibinfo {author} {\bibfnamefont {J.}~\bibnamefont
			{Skilling}},\ }\href@noop {} {\emph {\bibinfo {title} {Data Analysis: A
				Bayesian Tutorial}}},\ \bibinfo {edition} {2nd}\ ed.\ (\bibinfo  {publisher}
	{Oxford University Press},\ \bibinfo {address} {Oxford},\ \bibinfo {year}
	{2006})\BibitemShut {NoStop}%
	\bibitem [{\citenamefont {Ghanem}(2017)}]{Ghanem17}%
	\BibitemOpen
	\bibfield  {author} {\bibinfo {author} {\bibfnamefont {K.}~\bibnamefont
			{Ghanem}},\ }\emph {\bibinfo {title} {Stochastic Analytic Continuation: A
			Bayesian Approach}},\ \href {\doibase 10.18154/RWTH-2017-06704} {Ph.D.
		thesis},\ \bibinfo  {school} {RWTH Aachen University} (\bibinfo {year}
	{2017})\BibitemShut {NoStop}%
	\bibitem [{Spe()}]{Spektra}%
	\BibitemOpen
	\href@noop {} {\enquote {\bibinfo {title} {\url{www.spektra.app}},}\
	}\BibitemShut {NoStop}%
\end{thebibliography}
\end{document}